\documentclass[journal]{IEEEtran}

\usepackage{graphicx}
\usepackage{amsfonts,amsthm,bm}
\usepackage{tabularx}
\usepackage[makeroom]{cancel}
\usepackage{amssymb}
\usepackage{comment}
\usepackage{float}
\usepackage[dvipsnames]{xcolor}
\usepackage{ifpdf}
\usepackage{cite}
\usepackage{hyperref}
\usepackage{enumitem}
\usepackage{algorithm}
\usepackage{amsmath}
\usepackage[noend]{algpseudocode}
\usepackage{bbold}
\usepackage{multicol}
\usepackage{xparse}
\newtheorem{theorem}{Theorem}

\newtheorem{remark}{Remark}

\newtheorem{definition}{Definition}

\begin{document}

\title{Accelerated Probabilistic Power Flow {in Electrical Distribution Networks} via Model Order Reduction and Neumann Series Expansion}

\author{Samuel Chevalier,~\IEEEmembership{Member,~IEEE,}
        Luca Schenato,~\IEEEmembership{Fellow,~IEEE,}
        and Luca Daniel,~\IEEEmembership{Member,~IEEE}
\thanks{

This work was partially supported by the CARIPARO Visiting Programme 2018 ``HiPeR" n.9576.

S. Chevalier is with the Center for Electric Power and Energy, Department of Electrical Engineering, Technical University of Denmark (DTU), Lyngby, Denmark. Email: schev@elektro.dtu.dk.

{L. Schenato is with Department of Information Engineering, University of Padova, Padova, Italy. E-mail: schenato@dei.unipd.it}

{L. Daniel is with Department of EECS, Massachusetts Institute of Technology, Massachusetts, USA. E-mail: luca@mit.edu}}}

\maketitle

\begin{abstract}
This paper develops a computationally efficient algorithm which speeds up the probabilistic power flow (PPF) problem by exploiting the inherently low-rank nature of the voltage profile in electrical power distribution networks. The algorithm is accordingly termed the Accelerated-PPF (APPF), since it can accelerate ``any'' sampling-based PPF solver. As the APPF runs, it concurrently generates a low-dimensional subspace of orthonormalized solution vectors. This subspace is used to construct and update a reduced order model (ROM) of the full nonlinear system, resulting in a highly efficient simulation for future voltage profiles. When constructing and updating the subspace, the power flow problem must still be solved on the full nonlinear system. In order to {accelerate the computation of these solutions,} a Neumann expansion of a modified power flow Jacobian is implemented. Applicable when load {bus injections} are small, this Neumann expansion allows for a considerable speed up of Jacobian system solves during the standard Newton iterations. APPF test results, from experiments run on the full IEEE 8500-node test feeder, are finally presented.
\end{abstract}

\setlength{\belowdisplayskip}{6pt} 
\setlength{\abovedisplayskip}{6pt} 


\IEEEpeerreviewmaketitle


\section{Introduction}\label{Introduction}

\IEEEPARstart{T}{he} ongoing democratization of energy is causing a series of fundamental changes to electrical distribution grids. Distributed energy resources (DERs), such as Tesla powerwalls and rooftop photovoltaic systems, automated sensing devices equipped with telemetry capabilities, such as micro-Phasor Measurement Units ($\mu$-PMUs) and smart meters, and active loads, which are capable of reactively responding to real-time pricing signals, are all majorly disrupting the standard operating procedures of distribution networks~\cite{Boardman:2020,Kong:2018}. Consequentially, active management of the network’s resources is becoming critically important.

In order to properly operate and control these distribution grids, probabilistic forecasting of the network “state” (i.e., complex nodal voltages) is vitally important information, both for system operators and for the automated controllers embedded in the network. Performing such probabilistic forecasting in real-time can be computationally challenging, but it can also provide numerous operational benefits. The so-called probabilistic power flow (PPF)~\cite{Prusty:2017} maps uncertainties in the power injection space to corresponding performance uncertainties in the operational state; extensions of the PPF tool also incorporate network parameter uncertainty, although this paper deals exclusively with load uncertainty. The PPF has become an increasingly useful tool for system operators since its academic inception in 1974 ~\cite{Borkowska:1974,Allan:1974}. An excellent review on the topic is provided in~\cite{Prusty:2017}, although the state of the art has advanced considerably in recent years, due to the rapid improvement of advanced Uncertainty Quantification (UQ)\footnote{{UQ maps uncertainties from one space (e.g., the space of parameters, topologies, system inputs, etc.) to uncertainties in some output space.}} techniques{\cite{Soize:2017}}. 

Most PPF solvers fall into two general categories: simulation methods and analytical methods. Direct simulation approaches are typically referred to as Monte Carlo Simulations (MCS). These approaches attempt to directly build either the output distribution or quantities related to such distribution (i.e., statistical moments) through copious sampling and simulation of the underlying black box power flow solver. 

The analytical methods typically use mathematical simplifications and expansions in order to alleviate the computational burden associated with the direct simulation methods. For example, \cite{Allan:1974} uses a convolution of random variables in order to infer the output probability density function (PDF) of power flow solutions. The stochastic response surface method (SRSM) was first applied to the PPF in~\cite{Ren:2016}, where polynomial chaos expansion (PCE) is used to construct statistically equivalent output voltage distributions. Since then, the application of PCE to the PPF problem has seen a variety of improvements. In~\cite{Ni:2017}, optimal truncation and degree selection of the PC{E} series is considered, and nonlinear correlation of {random variables} is dealt with; in~\cite{Gruosso:2020}, so-called Stochastic Testing from~\cite{Zhang:2013} is applied to the generalized PCE in the context of time varying loads. Since PCE can suffer from the curse of dimensionality,~\cite{Sheng:2019} proposed the use of the low rank approximation (LRA), where the polynomial basis coefficient count grows linearly rather than exponentially. The cumulant~\cite{Zhang:2004,Fan:2012} and point estimate~\cite{Su:2005,Delgado:2014} methods are other relatively older, but still popular, analytical PPF methods. Both, though, must be augmented with series expansions in order for surrogate output PDFs to be constructed~\cite{Ren:2016}.

\begin{figure}
    \centering
    \includegraphics[width=1\columnwidth]{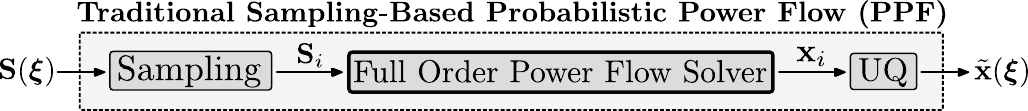}
    \caption{{Shown is a traditional sampling-based PPF procedure, where a random variable ${\bm \xi}$ parameterizes a load distribution ${\bf S}({\bm \xi})$. This distribution is sampled, and the resulting load profile ${\bf S}_i$ is passed to a power flow solver. Finally, the power flow solutions ${\bf x}_i$ are passed to an Uncertainty Quantification (UQ) tool which generates a surrogate {(i.e., representative)} voltage profile {distribution $\tilde{\bf x}({\bm \xi})$}. In this paper, we develop methods which speed up the power flow solver at the center of this routine, thus accelerating the entire PPF routine.}}
    \label{fig:TPPF_Routine}
\end{figure}

{Most all of the leading approaches in the literature, leveraging either simulation or analytical methods, are so-called sampling-based (also sometimes called \textit{non-intrusive}). This means they sample from an uncertainty set, pass the sample to a power flow solver, and then (optionally) apply some sort of UQ tool. An overview of this methodology is depicted in Fig.~\ref{fig:TPPF_Routine}.} There are two key aspects that determine the overall speed and efficiency of such sampling-based methods: the way they choose the samples where the system is to be solved; and the time it takes for each chosen sample to be solved. There are many contributions in the literature on improving the choice of the samples. For instance, in~\cite{Huang:2011}, importance sampling is applied to the probabilistic optimal power flow problem. Latin hypercube and Latin supercube sampling are employed in~\cite{Yu:2009} and~\cite{Hajian:2013}, respectively. These advanced sampling approaches attempt to limit the number of simulations needed to produce output PDFs of sufficiently high-fidelity. While these methods present some distinct advantages, sampling-based approaches tend to still be quite computationally heavy.

On the contrary, contributions which seek to speed up the individual ``black box" power flow solutions for a sampling-based PPF solver, especially in the context of distribution networks, have seldom been published, despite the fact that the power flow problem itself has been studied for many decades{\cite{Yang:2020}}. Ostensibly, solving power flow is not nearly as computationally burdensome as modern neural network training type problems. In a massive network, though, with three unbalanced phases, the problem can grow quite large, with tens of thousand of variables, and the numerics can become poorly conditioned. When it is desirable to solve thousands of power flows in a short time period, using only local computational resources, rapid system {solutions} can be highly attractive.

An essential tool for speeding up numerical solutions of physical systems is Model Order Reduction (MOR). {While MOR methodologies are often associated with \textit{dynamical} systems, they are also frequently used to simplify the computational burden of solving \textit{static} systems (e.g., solving large linear systems using projection approaches~\cite{El-Moselhy:2010})}. See~\cite{Benner:2015} for an excellent review of projection-based MOR techniques. More specifically,~\cite{El-Moselhy:2010} proposes a procedure which \textit{dynamically} constructs the projection subspace as an external solver runs. This procedure is highly applicable to the PPF problem. Intuitively, PPF subspace construction can be achieved by exploiting the practical observation that the vast majority of power flow solutions for the millions of different samples actually live within a small subspace (i.e., 10-50 dimensions). From an implementation perspective, this subspace is progressively uncovered from sequentially obtained solutions. In this paper, the subspace is concurrently used in a projection framework to construct and update a ROM; the highly-compact size of this ROM results in orders of magnitude faster solutions for the majority of the remaining samples. 

In this paper, we develop methods which alleviate the computational burden associated with applying PPF to distribution networks containing limited load uncertainty. Specifically, we present a series of methods which effectively speed up each sample solve (i.e., the solution of each load profile sample) for any sampling-based method performing PPF estimation. In the context of Fig.~\ref{fig:TPPF_Routine}, we are speeding up the power flow solver at the heart of the PPF routine. {The MOR methodology we use to accomplish this task, however, is not meant to be used as a way to solve a single arbitrary power flow problem; this is because (i) our methodology requires a nontrivial amount of computational overhead in order to progressively build up the associated ROM, and (ii) the resulting ROM may not generalize well to arbitrary power flow problems which are ``out of sample". Thus, our procedure is designed to be used within the particular confines of the PPF problem. The few other recent works which have sought to speed up power flows in the PPF have focused on computational acceleration (e.g., GPU-accelerated parallelization~\cite{Zhou:2018} and cloud-based computing strategies~\cite{Wang:2019Cloud}), machine learning prediction~\cite{Yang:2020}, and, more historically, the DC power flow approximation~\cite{Borkowska:1974},  which introduces unavoidable error. However, to the authors' knowledge, no other work has converted the exact AC nonlinear power flow equations into an equivalent reduced (or surrogate) model which is useful for directly solving samples from the probabilistic power flow problem.}

In this work, our methods were specifically developed for the unique characteristics of distribution networks. Otherwise, our work is completely agnostic to both the type of UQ methodology used to characterize the output distributions (i.e., PCE, Stochastic Collocation, etc.), and to the way samples are chosen (i.e., Monte Carlo, Important Sampling, etc.). We therefore refer to our routine as the \textit{Accelerated}-PPF (APPF) routine, since it can be used to speed up \textit{any} sampling-based PPF solver. The primary contributions of this paper follow:

\begin{enumerate}

\item We develop a fast power flow Jacobian solution technique which leverages a Neumann series approximation.

\item We use a dynamically expanding subspace to construct a ROM of the system which solves rapidly.

\item We combine the Jacobian solution and the reduced model solver to construct a routine which can rapidly solve many sequential power flow problems.

\end{enumerate}

The remainder of this paper is structured as follows. In Section \ref{eq: tech_back}, the power flow problem and the Neumann series expansion are reviewed. In Section \ref{Sec:NPFS}, a Neumann based power flow solver is derived, and in Section \ref{Sec:APPF}, a ROM with a dynamic subspace expansion routine is proposed. Here, the full APPF routine is also proposed. In section \ref{Sec:Results}, test results are provided from the 8500{-node distribution} network.


\section{Technical Background}\label{eq: tech_back}
In this section, we outline a standard network model and review the power flow problem. We then summarize the common numerical technique used to solve power flow. Finally, we recall the Neumann series expansion of a perturbed matrix, and we review a model order reduction based on Galerkin testing combined with dynamic expansion of its projection subspace.

\begin{remark}\label{def: linear_sys_solve}
In this paper, all notations of the type $A^{-1}{\bf b}$ (where $A$ is any given matrix and ${\bf b}$ is any given vector), should always be interpreted as performing a ``linear system solve" of system $A{\bf x}={\bf b}$  for some appropriately defined vector ${\bf x}$ (as opposed to inefficiently performing the computation of matrix inverse $A^{-1}$, followed by a matrix-vector product).
\end{remark}

\subsection{Standard Network Model Statement}
Our power flow methodology will be derived for a single-phase network and then extended to a three-phase network. In defining this network, we assume there is one unique bus which represents the distribution network {substation} (i.e., point of common coupling). When defining the model for a single-phase network, researchers often denote its graph as $G({\mathcal V},\bar{\mathcal E})$, with edge set $\bar{\mathcal E}$, $|\bar{\mathcal E}|=m$, vertex set $\mathcal{V}$, $|\mathcal{V}|=n$, and signed nodal incidence matrix ${\bar E}\in{\mathbb R}^{m\times n}$. 
\begin{remark}
For notational clarity, when defining variables which \textit{include} the {substation} node, an overline will be used. For instance, ${\bar E}$ contains all network nodes, including the {substation}.
\end{remark}
\noindent The nodal admittance (``Y-bus") matrix takes the form
\begin{align}\label{eq: Yb}
{\bar Y}_{b}={\bar E}^{\top}Y_{l}{\bar E}+{\bar Y}_{s},
\end{align}
where $Y_{l}\in {\mathbb C}^{m\times m}$ and ${\bar Y}_{s}\in {\mathbb C}^{n\times n}$ are the  diagonal line and shunt admittance matrices, respectively. In this network, we define ${\bar{\bf V}}e^{j{\bar{\bm \theta}}}\in {\mathbb C}^{n}$ and $\bar{\bf I}e^{j\bar{\bm \phi}} \in {\mathbb C}^{n}$ as the nodal voltage and nodal current injection phasor vectors, respectively, where ${\bar{\bf V}},\bar{\bm \theta},\bar{\bf I},\bar{\bm \phi}\in{\mathbb R}^{n}$. These vectors satisfy $\bar{\bf I}e^{j\bar{\bm \phi}} = {\bar Y}_b{\bar{\bf V}}e^{j{\bar{\bm \theta}}}$.
\subsection{The Power Flow Problem}
In this network, the \textit{deterministic} power flow problem seeks to determine the nodal voltage phasors which satisfy a set of nonlinear power flow equations. In polar form, the active and reactive power flow equations~\cite{Glover:2012} at node $i$ are written as
\begin{subequations}\label{eq: PF_eqs}
\begin{align}
P_{i} & =\mathrm{V}_{i}\sum_{k\in {\mathcal V}}\mathrm{V}_{k}\left(G_{ik}\cos(\theta_{ik})+B_{ik}\sin(\theta_{ik})\right)\\
Q_{i} & =\mathrm{V}_{i}\sum_{k\in {\mathcal V}}\mathrm{V}_{k}\left(G_{ik}\sin(\theta_{ik})-B_{ik}\cos(\theta_{ik})\right),
\end{align}
\end{subequations}
where $B_{ik}={\rm Im}\{\bar{Y}_{b}^{(ik)}\}$ and $G_{ik}={\rm Re}\{\bar{Y}_{b}^{(ik)}\}$ are susceptance and conductance values, respectively. Nonlinear system (\ref{eq: PF_eqs}) may be extended to include a full network of nodes and compactly written as $\bar{\bf S}=\bar{\bf s}(\bar{\bf x})$, where the vector of all nodal power injections and the vector of voltage magnitude and phase angle variables are, respectively,
\begin{align}
\bar{\bf S}&=\left[\bar{\bf P}^{\top},\,\bar{\bf{Q}}^{\top}\right]^{\top} \in {\mathbb R}^{2n}\\
\bar{\bf x}&=\left[\bar{\bf V}^{\top},\,\bar{\bm{\theta}}^{\top}\right]^{\top} \in {\mathbb R}^{2n}.
\end{align}
\begin{definition} A \textbf{power flow solution} is any vector $\bar{\bf x}$ satisfying
\begin{align}\label{eq: PF_sx=S}
\{\bar{\bf x}\in\mathbb{R}^{2n}\;|\;\Vert\bar{\bf s}(\bar{\bf x})-\bar{\bf S}\Vert<\epsilon\}.
\end{align}
\end{definition}
\noindent In practice, a power flow solver attempts to minimize residual function $\bar{\bf g}(\bar{\bf x}) \equiv \bar{\bf s}(\bar{\bf x})-\bar{\bf S}$, which codifies the mismatch between the specified and the predicted nodal power injections. Since power injection is only specified at a subset of nodes (i.e., not the {substation}), the residual to be minimized reduces:
\begin{align}\label{eq: res_func_red}
{ {\bf g}}({\bf x})\equiv{\bf s}({\bf x})-{\bf S}. 
\end{align}
In this paper, ${\bf x}$ is equivalent to $\bar{\bf x}$, but with the {substation} voltage (magnitude and phase) deleted. Similarly, ${\bf s}(\cdot)$ and ${\bf S}$ correspond to the power flow functions and injections, respectively, at all buses in the network \textit{except} the {substation}.

In the \textit{probabilistic} power flow (i.e., PPF) problem, there is uncertainty in the value of the loads in ${\bf S}$; this uncertainty is parameterized by some distribution vector ${\bm \xi}\in{\mathbb R}^{t}$ via ${\bf S}({\bm \xi})$, where generally $t \ll 2n$, depending on the number of uncertain loads. The PPF solver thus seeks to map the input distribution on the power injections to an output distribution on power flow solutions (i.e., voltage profiles) by solving ${\bf s}({\bf x})-{\bf S}(\xi)={\bf 0}$, as in~\cite{Sheng:2019}.

\subsection{Standard Numerical Solution Technique for Power Flow}
A power flow solver seeks to minimize the residual (\ref{eq: res_func_red}) of the nonlinear power flow equations. This system represents an equal number of equations and unknowns. Newton-Raphson iterations are most commonly used to solve this system:
\begin{align}
\label{eq: power_flow_it}
{\rm Solve}\!: { J}({\bf x}^{(i)})
\Delta{\bf x}^{(i)} &= - {\bf g}({\bf x}^{(i)}),\\
{\bf x}^{(i+1)}&\leftarrow{\bf x}^{(i)} + \Delta {\bf x}^{(i)},\nonumber
\end{align}
where $ { J}({\bf x}^{(i)})$ is the \textit{reduced} power flow Jacobian (RPFJ) matrix, which is typically constructed using the summation of explicit partial derivative terms~\cite{Glover:2012}. In~\cite{Bolognani:2015}, however, Bolognani and D\"{o}rfler propose a novel Jacobian structure:
\begin{align}\label{eq: RPFJ}
{J}({\bf x}^{(i)})=(\langle{\mathtt d}({{\bf I}}e^{-j{\bm{\phi}}})\rangle+\langle{\mathtt d}({{\bf V}}e^{j{\bm{\theta}}})\rangle{N}\langle{Y}_{b}\rangle)R({{\bf V}}e^{j{\bm{\theta}}}),
\end{align}
where ${\mathtt d}(\cdot)$ is the ``$\rm diag$" operator. The terms $R(\cdot)$, $N$, and $\langle \cdot \rangle$ are given in~\cite{Bolognani:2015} and contextually demonstrated in (\ref{eq: J_terms}):
\begin{subequations}\label{eq: J_terms}
\begin{align}\langle\mathtt{d}({{\bf I}}e^{-j{\boldsymbol{\phi}}})\rangle & =\left[\begin{array}{cc}
{\rm Re}\{\mathtt{d}({{\bf I}}e^{-j{\boldsymbol{\phi}}}\} & {\rm Im}\{\mathtt{d}({{\bf I}}e^{-j{\boldsymbol{\phi}}})\}\\
-{\rm Im}\{\mathtt{d}({{\bf I}}e^{-j{\boldsymbol{\phi}}})\} & {\rm Re}\{\mathtt{d}({{\bf I}}e^{-j{\boldsymbol{\phi}}})\}
\end{array}\right]\\
\langle\mathtt{d}({{\bf V}}e^{j{\boldsymbol{\theta}}})\rangle & =\left[\begin{array}{cc}
{\rm Re}\{\mathtt{d}({{\bf V}}e^{j{\boldsymbol{\theta}}})\} & -{\rm Im}\{\mathtt{d}({{\bf V}}e^{j{\boldsymbol{\theta}}})\}\\
{\rm Im}\{\mathtt{d}({{\bf V}}e^{j{\boldsymbol{\theta}}})\} & {\rm Re}\{\mathtt{d}({{\bf V}}e^{j{\boldsymbol{\theta}}})\}
\end{array}\right]\\
N\langle{Y}_{b}\rangle & =\left[\!\begin{array}{cc}
E^{\top}{Y}_{l}^{g}E & -E^{\top}{Y}_{l}^{b}E\!-\!{Y}_{s}^{b}\\
-E^{\top}{Y}_{l}^{b}E\!-\!{Y}_{s}^{b} & -E^{\top}{Y}_{l}^{g}E
\end{array}\!\right]\\
R({{\bf V}}e^{j{\boldsymbol{\theta}}}) & =\left[\begin{array}{cc}
\mathtt{d}(\cos({\boldsymbol{\theta}})) & -{\rm Im}\{\mathtt{d}({{\bf V}}e^{j{\boldsymbol{\theta}}})\}\\
\mathtt{d}(\sin({\boldsymbol{\theta}})) & {\rm Re}\{\mathtt{d}({{\bf V}}e^{j{\boldsymbol{\theta}}})\}
\end{array}\right],
\end{align}
\end{subequations}
where ${\bf V}e^{j\bm{\theta}}$, ${\bf I}e^{j\bm{\phi}}$ are appropriately reduced voltage, current vectors. The diagonal matrices ${Y}^b_{s}={\rm Im}\{{Y}_{s}\}$, ${Y}^g_{l} = {\rm Re}\{{Y}_{l}\}$, and ${Y}_{l}^{b} = {\rm Im}\{{Y}_{l}\}$ come from the ``reduced" Y-bus matrix:
\begin{align}\label{eq: R_Yb}
{Y}_{b}&={E}^{\top} {Y}_{l} {E}+{Y}_{s}.
\end{align}
Matrix $E$ comes from eliminating the root node column $c_1$ of the full incidence matrix: ${\bar E}=\left[\!\!\begin{array}{cc}
c_{1} \!\!&\!\! E \end{array}\!\!\right]$.

\subsection{Neumann Series Expansion of a Perturbed Matrix}
In this paper, we will make use of the following standard result.
Consider a linear system 
\begin{equation}
\label{eq:ParameterizedSystem}
(A+\epsilon D) {\bm x} = {\bf b}
\end{equation}
where $A$ is a constant matrix, $\epsilon D$ is a \textit{changing} perturbation matrix with some small scalar $\epsilon$, and ${\bf b}$ is 
a \textit{changing} right hand side. 
The $k^{\rm th}$ Neumann series expansion{\cite{Horn:1990}} of the matrix $(A+\epsilon D)^{-1}$ can be used to approximate the solution $\bm x$:
\begin{align}\label{eq: NS}
\bm{x} & \approx  \sum_{i=0}^{k}(-1)^{i}(A^{-1}\epsilon D)^{i}A^{-1}{\bf b} + \mathcal{O}(\epsilon^{k+1}).
\end{align}
The series (\ref{eq: NS}) converges when $\epsilon<1/\rho(A^{-1}D)$~\cite{Moselhy:2011}, where $\rho(\cdot)$ is the spectral radius operator. 
To efficiently compute (\ref{eq: NS}), one can first decompose $A$ into its $LU$ factors. Then, each instance of $A^{-1}$ can be implemented with efficient {solvers} by using the LU factors in standard forward-elimination and back-substitution routines. After initializing $\bm{z}^{(0)}=\bm{x}^{(0)}$ 
by solving $LU\bm{x}^{(0)}={\bf b}$,
one can then iterate until the desired accuracy is reached:
\begin{align}
\label{eq: NS_it}
{\rm Solve}\!: LU \bm{z}^{(i+1)} = \epsilon D\bm{z}^{(i)}\\
\!\!\!\ensuremath{\bm{x}^{(i+1)}\leftarrow\bm{x}^{(i)}\!+\!(-1)^{i}\bm{z}^{(i+1)}}.\nonumber
\end{align}
\begin{remark}
The complexity of a single iteration is  ${\mathcal O}(n^2)$ when $A$ and $D$ are dense, and it is ${\mathcal O}(n)$ when they are sparse.
\end{remark}

\subsection{Model Order Reduction}\label{ss_MOR_DSE}
A standard way to further speed up 
the solution of system (\ref{eq:ParameterizedSystem}) involves looking for a solution $\hat{{\bm x}}\in \mathbb R^q$ in a low dimensional subspace. One may represent the solution
\begin{align}
\label{eq: Vxh}
{\bm x}\approx V \hat{\bm x}
\end{align}
as a linear combination of columns of an orthonormal projection operator $V \in \mathbb R^{n\times q}, q \ll n$. The approximate solution can be obtained by efficiently solving the reduced system generated by a standard Galerkin testing~\cite{Zhang:2013}:
\begin{equation}
\label{eq:ReducedSystem}
(\hat{A}+\epsilon\hat{D}) \hat{\bm x} = \hat{\bf b},
\end{equation}
where $\hat{A}=V^{\top} A V\!\in\! \mathbb R^{q\times q}$, $\hat{D}=V^{\top} D V\!\in\! \mathbb R^{q\times q}$ and $\hat{\bf b} = V^{\top} {\bf b}\!\in \mathbb R^{q}$.
The literature on projection based model order reduction provides many options for constructing operator $V$.  Later in this paper, we will modify and make use of some of the techniques and theoretical results in~\cite{Daniel:2004} and in~\cite{El-Moselhy:2010} for dynamic update of parameterized reduced order models. {For convenience, these projection and update procedures are combined and summarized in Algorithm~\ref{alg:MORDU}. Clearly, these methods are applied to a \textit{linear} system in this algorithm. Our paper, however, will apply similar methods to the \textit{nonlinear} system of power flow equations.}


\begin{algorithm}
\caption{Model Order Reduction with Dynamic Update}\label{alg:MORDU}
{\small
\begin{algorithmic}[1]
\State ${\bm x}\leftarrow$ Solve full-order system (\ref{eq:ParameterizedSystem}), e.g., with Neumann (\ref{eq: NS_it})
\State $V\leftarrow {\bm x}/\Vert{\bm x}\Vert, \;\;
\hat{A}\leftarrow V^{\top}AV$
\For{each new $D$ and/or ${\bf b}$}
\State $\hat{D}\leftarrow V^{\top}DV, \;\;
\hat{\bf b}\leftarrow V^{\top}{\bf b}$
\State $\hat{\bm x}\leftarrow$ Solve Reduced System (\ref{eq:ReducedSystem})
\If {$||(A + \epsilon D) V\hat{\bm x} - {\bf b} ||>$ tolerance}
\State ${\bm x}\leftarrow$ Solve full-order system (\ref{eq:ParameterizedSystem})
\State ${\bm v}\leftarrow {\bm x} - VV^{\top}{\bm x}$
$\quad\quad\quad\,\triangleright$ remove projection into $V$
\State $V\leftarrow [V\;\; {\bm v}/\Vert {\bm v} \Vert]$ $\quad\quad\quad\,\triangleright$ extend subspace
\State $\hat{A}\leftarrow 
                \begin{bmatrix}
                \hat{A} & V^{\top}A{\bm v}\\
                {\bm v}^{\top}AV   & {\bm v}^{\top}A{\bm v}
                \end{bmatrix}$
$\triangleright$ update reduced model
\EndIf
\EndFor
\end{algorithmic}}
\end{algorithm}

\section{Computationally Efficient Power Flow Solution via Neumann Series Expansion}\label{Sec:NPFS}
The primary computational bottleneck in power flow is solving (\ref{eq: power_flow_it}). 
In this section, we show how to
solve system~(\ref{eq: power_flow_it}) efficiently in ${\mathcal O}(n)$ operations{\cite{Horn:1990}} via Neumann series expansion iterations~(\ref{eq: NS_it}). {That is, we exploit precomputed sparse LU factorizations in order to efficiently solve the linear system which is at the heart of the power flow problem.} Effectively, this is accomplished by putting Jacobian~(\ref{eq: RPFJ}) 
into framework~(\ref{eq:ParameterizedSystem}). We further propose practical convergence criteria. Finally, we outline the power flow algorithm, {we show its applicability to other popular power flow solvers,} and we offer extensions to three-phase systems.

\subsection{Neumann Series Expansion Applied to Power Flow}
We rewrite linear system (\ref{eq: power_flow_it}) as two sub-systems:
\begin{align}
\big(\langle{\mathtt d}(\tilde{{\bf I}}^{*})\rangle+\langle{\mathtt d}(\tilde{{\bf V}})\rangle{N}\langle{E}^{\top}Y_{l}{E}+{Y}_{s}\rangle\big){\bf y} & ={\bf b}\label{eq: Ay=b}\\
R(\tilde{{\bf V}})\Delta{\bf x} & ={\bf y},\label{eq: Rx=y}
\end{align}
where, for convenience, ${\bf b} = -{\bf g}({\bf x})$, $\tilde{{\bf V}}={\bf V}e^{j\bm{\theta}}$, $\tilde{{\bf I}}={\bf I}e^{j\bm{\phi}}$.
\begin{remark}
System (\ref{eq: Rx=y}) can be solved in ${\mathcal O}(n)$.
\end{remark}
\noindent Considering (\ref{eq: Ay=b}), we multiply both sides by $\langle{\mathtt d}(\tilde{{\bf V}})\rangle^{-1}$ to yield
\begin{align}
\label{eq: DpLU=b}
\!\!\!\big(
\underbrace{{N}\langle{E}^{\top}Y_{l}{E}+{Y}_{s}\rangle}_{\mathcal{L}\mathcal{U}}\!
+
\!\underbrace{\langle{\mathtt d}(\tilde{{\bf V}})\rangle^{-1}\!\langle{\mathtt d}(\tilde{{\bf I}}^{*})\rangle}_{\mathcal{D}}
\big){\bf y}
=
\underbrace{\langle{\mathtt d}(\tilde{{\bf V}})\rangle^{-1}{\bf b}}_{{\bm b}}\!.
\end{align}

\begin{remark}\label{Re: LDL}
Matrix ${N}\langle{E}^{\top}Y_{l}{E}+{Y}_{s}\rangle$ will always be symmetric in a per-unitized network. Therefore, we perform an $LDL^{\top}$ decomposition, and assign ${\mathcal L} = L$ and ${\mathcal U} = DL^{\top}$. There might also be orthogonal permutation matrix ${\mathcal P}$, such that
\begin{align}\label{eq: LDL}
\mathcal{P}^{\top}({N}\langle{E}^{\top}Y_{l}{E}+{Y}_{s}\rangle)\mathcal{P}=\mathcal{L}\mathcal{U}.
\end{align}
\end{remark}

\begin{remark}
{Since both $\mathcal D$ and ${\bm b}$ in (\ref{eq: DpLU=b}) represent solutions of ultra-sparse linear systems (i.e.,  $\mathtt{d}(\tilde{{\bf V}})\rangle\mathcal{D}=\langle\mathtt{d}(\tilde{{\bf I}}^{*})\rangle$ and $\langle\mathtt{d}(\tilde{{\bf V}})\rangle\bm{b}={\bf b}$),} they can both be computed in ${\mathcal O}(n)$.
\end{remark}

\noindent
In summary,
in ${\mathcal O}(n)$ operations,
we can express 
(\ref{eq: power_flow_it}) as 
\begin{align}
\label{eq: D+LU}
\left(\mathcal{L}\mathcal{U}+\mathcal{D}\right){\bf y}={{\bm b}},
\end{align} 
via (\ref{eq: DpLU=b}), where
$\mathcal L$, $\mathcal U$ are lower and upper triangular matrices respectively. Since the $\mathcal{L}\mathcal{U}$ factors are functions of network admittance parameters and topology, their values are constant at each iteration. When considering the most efficient way to solve (\ref{eq: D+LU}), we note that matrix ${\mathcal D}$ is composed of load currents scaled by nodal voltages. In a distribution network, these are typically very small compared to the ${\mathcal L}{\mathcal U}$ matrix elements. {Since $\Vert\mathcal{D}\Vert\ll\Vert\mathcal{LU}\Vert$,} we therefore consider ${\mathcal D}$ as a perturbation applied to matrix $\mathcal{L}\mathcal{U}$, { just as $\epsilon D$ is a perturbation in (\ref{eq:ParameterizedSystem}).} Accordingly, we approximate the solution of (\ref{eq: D+LU}) by applying the Neumann series iteration from (\ref{eq: NS_it}){:
\begin{align}\label{eq: NS_PF}
\bf{y} & \approx \sum_{i=0}^{k}(-1)^{i}(({\mathcal LU})^{-1} \mathcal{D})^{i}({\mathcal LU})^{-1}{\bm b}.
\end{align}}
{Larger values of $k$ will generally yield more accurate approximations. In practice, we observed that a $3^{\rm rd}$ order Neumann series (i.e., $k=3$) yielded approximations with less than 1\% error on the 8500 node distribution grid. We note, though, that this represents error in a single iterative Newton step, and not the error in the final power flow solution (which can be solved to much tighter degrees of accuracy).}

\subsection{Practical Neumann Series Convergence Criteria}
To justify this Neumann series application, the following theorem presents practical convergence condition (\ref{eq: suff_cond}). This theorem deals with complex, rather than real, coordinates. Accordingly, we define $\mathcal{D}_c ={\mathtt d}(\tilde{{\bf V}})^{-1}{\mathtt d}(\tilde{{\bf I}}^{*})$ and $\mathcal{L}_c\mathcal{U}_c ={E}^{\top}Y_{l}{E}$.
\begin{theorem}\label{Theorem: NS}
 Matrix inverse $(\mathcal{L}_{c}\mathcal{U}_{c}+\mathcal{D}_{c})^{-1}$ can be approximated by a Neumann series iteration, as in (\ref{eq: NS_it}), if
\begin{align}\label{eq: suff_cond}
\rho({E}^{\top}Y_{l}{E})>{\rm max}\{|\tilde{{\bf I}}|\}.
\end{align}
\end{theorem}
See Appendix \ref{AppA} for the proof. If the largest load current magnitude is safely below $\rho({E}^{\top}Y_{l}{E})$, then the Neumann series will converge. Since load currents are typically less than 1 p.u. and network admittance values are typically much larger, this condition is usually satisfied by a large margin. For example, in the IEEE 123-bus network~\cite{Schneider:2018}, $\Vert\mathcal{L}\mathcal{U}\Vert/\Vert\mathcal{D}\Vert\approx4\!\times\!10^4$.

\subsection{Solving Power Flow via Neumann Expansion}
The full power flow procedure, outlined in Algorithm \ref{alg:NPFS}, iterates until Newton converges according to some tolerance on the residual injection at each bus. This solver is termed the Neumann Power Flow Solver (\textbf{NPFS}). At several points, this power flow solver calls Algorithm \ref{alg:FEBS}, which solves the pre-factored network admittance matrix using forward-elimination and backward-substitution.

\begin{algorithm}
\caption{Neumann Power Flow Solver (NPFS)}\label{alg:NPFS}

{\small \textbf{Require:} Matrix factors $\mathcal L$, $\mathcal U$, $\mathcal P$ from (\ref{eq: LDL}); specified power injections ${ {\bf S}}$; initial voltage guess ${{\bf x}}_0$; reduced power flow function ${\bf s}(\cdot)$

\textbf{Ensure:} Solution $ {\bf x}$ satisfies ${{\bf s}}({\bf x})  \approx {{\bf S}}$

\begin{algorithmic}[1]

\Function{${\bf x} \leftarrow\,$NPFS}{${\mathcal L},{\mathcal U},{\mathcal P},{{\bf S}},{\bf x}_0$}

\State $k\leftarrow 0$

\State $ {\bf b} \leftarrow {{\bf S}}-{{\bf s}}({\bf x}_k)$

\While{$\Vert{\bf b}\Vert_\infty > $ tolerance $\epsilon_N$}

\State Construct $\langle{\mathtt d}(\tilde{{\bf V}})\rangle$ and $\langle{\mathtt d}(\tilde{{\bf I}}^{*})\rangle$ from ${\bf x}_k$

\State ${\bm b} \leftarrow$ Solve:  $\langle{\mathtt d}(\tilde{{\bf V}})\rangle{\bm b}={\bf b}$

\State ${\mathcal D} \leftarrow$ Solve: $\langle{\mathtt d}(\tilde{{\bf V}})\rangle{\mathcal D}=\langle{\mathtt d}(\tilde{{\bf I}}^{*})\rangle$

\State $i\leftarrow 0$


\State ${\bf y}^{(0)} = {\bf z} \leftarrow\Call{FEBS}{{\mathcal L},{\mathcal U},{\mathcal P},{\bm b}}$

\For{desired number of NS iterations}

\State $i \leftarrow i + 1$

\State ${\bf z} \leftarrow \Call{FEBS}{{\mathcal L},{\mathcal U},{\mathcal P},{\mathcal D}{\bf z}}$


\State ${\bf y}^{(i+1)} \leftarrow {\bf y}^{(i)} + (-1)^i {\bf z}$

\EndFor {\bf end}

\State $\Delta{\bf x}_k \leftarrow$ Solve: $R(\tilde{{\bf V}})\Delta{\bf x}_k ={\bf y}$

\State ${\bf x}_{k+1} \leftarrow {\bf x}_{k} - \Delta{\bf x}_k$

\State ${\bf b} \leftarrow {{\bf S}}-{{\bf s}}({\bf x}_k)$

\State $k \leftarrow k + 1$

\EndWhile {\bf end}

\State \Return ${\bf x}$

\EndFunction

\end{algorithmic}}
\end{algorithm}

\begin{algorithm}
\caption{Forward-Elimination Back-Substitution (FEBS)}\label{alg:FEBS}

 {\small \textbf{Require:}
 Matrix factors $\mathcal L$, $\mathcal U$, $\mathcal P$; RHS vector $\bf b$
 
\textbf{Ensure:} Returned $\bf x$ satisfies $\mathcal{P}\mathcal{L}\mathcal{U}\mathcal{P}^{\top}{\bf x}={\bf b}$

\begin{algorithmic}[1]

\Function{${\bf x} \leftarrow$FEBS}{${\mathcal L}, {\mathcal U}, {\mathcal P}, {\bf b}$}

\State $ {\bf z} \leftarrow$ solve (${\mathcal L} {\bf z} = {\mathcal P}^{\top} {\bf b}$) with Forward-Elimination

\State $ {\bf y} \leftarrow$ solve (${\mathcal U} {\bf y}={\bf z}$) with Back-Substitution

\State ${\bf x} \leftarrow {\mathcal P}{\bf y}$

\State \Return ${\bf x}$

\EndFunction

\end{algorithmic}}
\end{algorithm}

{\subsection{Applications to Other Power Flow Solvers}
There exist a variety of other power flow solvers which are faster than the considered full-order Newton-Raphson routine of (\ref{eq: power_flow_it}). Two of the most popular are the Fast-Decoupled Power Flow~\cite{Glover:2012}, and the quasi-Newton methods which utilize partial Jacobian updates (PJU)~\cite{Leon:2002}. In the sequel, we show that the NPFS routine can be applied to both of these methods.
\subsubsection{Fast-Decoupled Power Flow} Fast-Decoupled methods neglect Jacobian sub-matrices $J_{{\bf P},{\bf V}}$ and $J_{{\bf Q},{\bm \theta}}$, thus decreasing computational complexity. With the assumption that phase angle differences are small ($\sin(\theta)\approx 0$) and $J_{{\bf P},{\bf V}}\approx J_{{\bf Q},{\bm \theta}}\approx {\bf 0}$, the full system (\ref{eq: DpLU=b}) can be decoupled into two subsystems
\begin{subequations}\label{eq: LU_decouple}
\begin{align}
(\mathcal{L}_{1}\mathcal{U}_{1}+\mathcal{D}_{1}){\bf y}_{1} & =\bm{b}_{1}\\
(\mathcal{L}_{2}\mathcal{U}_{2}+\mathcal{D}_{2}){\bf y}_{2} & =\bm{b}_{2},
\end{align}
\end{subequations}
which can be iterated upon individually. {The authors in~\cite{Lee:2003} also use a Neumann series approximation in order to alleviate the computational complexity of a decoupled power flow problem; however, their implementation requires a new LU factorization of the sub-Jacobians at each iteration, and the convergence properties of the algorithm are never investigated.}
\subsubsection{Quasi-Newton Power Flow} Quasi-Newton methods do not perform full updates of the Jacobian matrix. If this is desirable, the perturbation matrix $\mathcal D$ of (\ref{eq: DpLU=b}), which is a function of the state iteration ${\bf x}^{(j)}$, can be updated fully, partially, or not at all. Denoting a partial update of $\mathcal D$ as $\check{\mathcal D}({\bf x}^{(j)})$, the Neumann approximation (\ref{eq: NS_PF}) is updated to 
\begin{align}
{\bf y} \approx \sum_{i=0}^{k}(-1)^{i}(({\mathcal LU})^{-1} {\check{\mathcal{D}}}({\bf x}^{(j)}))^{i}({\mathcal LU})^{-1}{\bm b}.
\end{align}}

\subsection{Extension to Unbalanced Three-Phase Networks}
The numerical routine proposed in Algorithm \ref{alg:NPFS} can be readily extended to an unbalanced, three-phase network. In these networks, each element must be described by \begin{align}\label{eq: 3Phase_primY}
\left[\begin{array}{c}
\tilde{\mathsf{I}}_{ij}\\
\tilde{\mathsf{I}}_{ji}
\end{array}\right]=\left[\begin{array}{cc}
\mathsf{Y}_{ii} & \mathsf{Y}_{ji}\\
\mathsf{Y}_{ij} & \mathsf{Y}_{jj}
\end{array}\right]\left[\begin{array}{c}
\tilde{\mathsf{V}}_{i}\\
\tilde{\mathsf{V}}_{j}
\end{array}\right],
\end{align}
where the stated admittance matrix is $6\times6$ and $\tilde{\mathsf{I}}_{ij},\,\tilde{\mathsf{V}}_{i}\in\mathbb{C}^{3\times1}$. This formulation assumes properly grounded neutral phases which are eliminated via Kron reduction. If one or more of the phases do not exist (i.e., ``disjoint" phase extensions), then the corresponding columns of the admittance matrix (\ref{eq: 3Phase_primY}) are removed. The three phase nodal admittance matrix associated with the full network is constructed by properly placing and summing the individual primitive admittance matrices. This nodal admittance matrix, including shunts, is written as $\mathsf{Y}_b$. For more details on formulating the 3-phase admittance matrix, see~\cite{Dugan:2012}. Using this matrix, the three-phase reduced power flow Jacobian can be directly constructed as
\begin{align}
{\mathsf{J}} & =\big(\langle{\mathtt d}(\tilde{\mathsf{{I}}}^{*})\rangle+\langle{\mathtt d}(\tilde{\mathsf{{V}}})\rangle{\mathsf{N}}\langle{\mathsf Y}_b\rangle\big)R(\tilde{\mathsf{{V}}}).\label{eq: Jac_3}
\end{align}
Any undefined terms can be inferred from (\ref{eq: RPFJ}), with the guiding assumption that both current and voltage vectors are concatenated $n$-phase vectors, where $n=1,\,2,\,3$ depending on the nature of the disjoint phase extensions. 
\begin{remark}
The admittance ${\mathsf{N}}\langle{\mathsf Y}_b\rangle$ from (\ref{eq: Jac_3}) will necessarily be symmetric. Accordingly, the LDL decomposition proposed in Remark \ref{Re: LDL} will be a valid factorization for this matrix.
\end{remark}
Other three-phase structures can be constructed analogously. Using these updated expressions, the solution to the three-phase power flow can proceed as outlined in Alg. \ref{alg:NPFS}.

\section{Reduced Order Modeling of the Nonlinear Power Flow for Probabilistic Power Flow}\label{Sec:APPF}
In this section, we first motivate the low-rank nature of power flow solutions in the PPF problem. Next, we leverage a projection subspace in order to perform model order reduction on the full power flow problem, and we show how Newton iterations can be used to solve the associated over-determined nonlinear system. Finally, we combine our proposed model order reduction method with our Neumann series based power flow algorithm in order to quickly solve for PPF solutions.

\subsection{Model Order Reduction of the Power Flow Problem}
A PPF solver considers the probable loading levels of a distribution network over some probabilistic horizon. The corresponding set of operating points is inherently low-rank. Anchored by tightly regulated feeder voltages, the network voltage profile ${\bf x}\in{\mathbb R}^{n}$ typically lives in a fairly low-dimensional subspace characterized by $V\!\in\!{\mathbb R}^{n \times q}$ from (\ref{eq: Vxh}). When $V$ is populated with a sufficient number of appropriately chosen dominant basis vectors, the low order vector $\hat{{\bf x}}$ can represent the full-order state with a high degree of accuracy.


To exploit the usefulness of subspace $V$ beyond the Galerkin projection of (\ref{eq:ReducedSystem}), we note that the power flow equations become quadratic when written in Cartesian coordinates $V_{\rm r}$, $V_{\rm i}$. We can therefore express the reduced residual function (\ref{eq: res_func_red}) as an \textit{exact} second order Taylor series expansion. Writing the reduced voltage vector in Cartesian coordinates as ${\bf x}_c$,
\begin{align}\label{eq: resid_func}
{\bf g}(\delta{\bf x}_{c})={\bf s}({\bf x}_{c0})+J_{c0}\delta{\bf x}_{c}+\tfrac{1}{2}H_{c}\left(\delta{\bf x}_{c}\otimes\delta{\bf x}_{c}\right)-{\bf S},
\end{align}
where $\delta{\bf x}_{c}={\bf x}_{c}-{\bf x}_{c0}$ is a perturbation from some nominal operating point ${\bf x}_{c0}$ and $\otimes$ is the Kronecker product. Refer to Appendix \ref{AppB}
for the construction of the associated Jacobian $J_{c0}\equiv J_{c}({\bf x}_{c0})$ and Hessian $H_c$ matrices.
Leveraging ${\bf x}_c\approx V\hat{{\bf x}}_c$, as in (\ref{eq: Vxh}), we note that $V\hat{{\bf x}}_{c}=V\hat{{\bf x}}_{c0}+V\delta\hat{{\bf x}}_{c}$. Substituting $V\delta\hat{{\bf x}}_{c}\approx\delta{\bf x}_{c}$ into (\ref{eq: resid_func}),
\begin{align}
\!\!{\bf g}(V\delta\hat{{\bf x}}_{c}) \!={\bf s}_{0}\!+\!J_{c0}V\delta\hat{{\bf x}}_{c}\!+\!\tfrac{1}{2}H_{c}(V\!\!\otimes\!\! V)(\delta\hat{{\bf x}}_{c}\!\otimes\!\delta\hat{{\bf x}}_{c})\!-\!{\bf S},\label{eq: Kron_sep}
\end{align}
where ${\bf s}_0={\bf s}({\bf x}_0)$ and Kronecker products in (\ref{eq: Kron_sep}) have been separated~\cite{Dong:2003}. Minimizing residual ${\bf g}(V\delta\hat{{\bf x}}_{c})$ in a least squares sense is the minimization of an overdetermined nonlinear system. Applying a Newton-like method to the associated least squares minimization yields the iterative routine
\begin{align}\label{eq: step_xc_Newton}
\delta\hat{{\bf x}}_{c}^{(i+1)} & =\delta\hat{{\bf x}}_{c}^{(i)}-{\hat G}^{-1}{\hat{\bf g}}(\delta\hat{{\bf x}}_{c}^{(i)}),
\end{align}
where $\hat{G}$ is the result of a modified Galerkin projection: $\hat{G}=(J_{c0}V)^{\top}\!J_{c0}V$. The details behind the formulation of (\ref{eq: step_xc_Newton}) can be found in Appendix \ref{App_min_r}. {We note that the iterative routine (\ref{eq: step_xc_Newton}) solves a \textit{nonlinear} reduced order model.} Once converged, $\delta\hat{{\bf x}}_{c}^{(i)}$ will be equal to the reduced state deviation which is the least squares minimizer of (\ref{eq: Kron_sep}). Notably, the residual of ${\hat{\bf g}}(\delta\hat{{\bf x}}_{c}^{(i)})$ (see appendix) and the steps of (\ref{eq: step_xc_Newton}) can both be computed very quickly, since the reduced system is extremely small ($q\times q$). A key property of this low-dimensional system is that (\ref{eq: step_xc_Newton}) reliably converges even though $\hat G$ is not updated between iterations; as will be shown, it is only updated when the basis $V$ dynamically expands. The procedure associated with {solving the} ROM system (\ref{eq: step_xc_Newton}) is given in Algorithm \ref{alg:RMS} and is termed the Reduced Model Solver (\textbf{RMS}).

\begin{algorithm}
\caption{Reduced Model Solver (RMS)}\label{alg:RMS}
{\small

\begin{algorithmic}[1]

\Function{$[\delta\hat {\bf x}_c,{\bf x}]\leftarrow\,$RMS}{$V,{\hat{\bf s}}_0,{\hat G},{\hat H},\hat{\bf x}_{c0},\delta\hat{\bf x}_{c}^{(1)},{\hat {\bf S}}$}

\State $\hat{{\bf g}}\leftarrow\hat{{\bf s}}_{0}+\hat{G}\delta\hat{\bf x}_{c}^{(1)}+\tfrac{1}{2}\hat{H}(\delta\hat{\bf x}_{c}^{(1)}\otimes\delta\hat{\bf x}_{c}^{(1)})-\hat{{\bf S}}$

\State $i \leftarrow 1$

\While{$\Vert \hat{\bf g}\Vert_{\infty}>$ tolerance ${\hat \epsilon}_N$}

\State $\delta\hat{{\bf x}}_{c}^{(i+1)} \leftarrow \delta\hat{{\bf x}}_{c}^{(i)}-{\hat G}^{-1}{\hat{\bf g}}(\delta\hat{{\bf x}}_{c}^{(i)})$

\State $i \leftarrow i+1$

\State $\hat{{\bf g}} \leftarrow \hat{{\bf s}}_{0}+\hat{G}\delta\hat{{\bf x}}_{c}^{(i)}+\tfrac{1}{2}\hat{H}(\delta\hat{{\bf x}}_{c}^{(i)}\otimes\delta\hat{{\bf x}}_{c}^{(i)})-\hat{{\bf S}}$

\EndWhile \textbf{end}

\State \Return $\delta\hat{\bf x}_c \leftarrow \delta\hat{\bf x}_{c}^{(i)}$, ${\bf x} \leftarrow$ Cartesian-to-Polar$\{V (\delta\hat{\bf x}_c\!+\!\hat{\bf x}_{c0})\}$

\EndFunction
  
\end{algorithmic}}
\end{algorithm}

Failure of convergence of Alg. \ref{alg:RMS} has not been witnessed by the authors. Such potential failure, though, would always be detected by the APPF algorithm, as shown in Fig. \ref{fig:PPF}, in which case the full-order model will be used to solve the system.

\subsection{Dynamic Subspace Expansion}
The quality of the RMS solution is a function of how effectively the subspace $V$ has been ``filled out". In order to expand $V$, we assume we have an emerging sequence of valid power flow solutions in Cartesian coordinates. We then leverage the dynamic subspace expansion technique characterized by line 9 in Algorithm \ref{alg:MORDU}. In this way, the basis $V \in \mathbb R^{n\times q}$ is \textit{dynamically} constructed as an outer loop PPF solver runs {and $q$ grows in size. We note that the ultimate value of $q$ is not known or chosen a priori; instead, it is allowed to grow whenever $V$ needs to add a new column to expand its column space.}

Such dynamic updating is computationally cheap, and it ensures that $V$ only contains subspace vectors which are useful for solving a particular PPF problem. Indeed, $V$ growing too large will slow the iterative scheme (\ref{eq: step_xc_Newton}) down considerably. As $V$ grows in size, though, the quality of the RMS results will improve. The expansion procedure is outlined in Algorithm \ref{alg:DSE}, which also includes updates of $\hat J$, $\hat G$, $\hat H$, $\delta\hat{\bf x}_c$ and ${\hat {\bf s}}_0$. Notably, the ``${\rm sort}(\cdot)$" function ensures that $\hat H$ is properly ordered, so it correctly interacts with Kronecker product $\delta\hat{{\bf x}}_{c}\otimes\delta\hat{{\bf x}}_{c}$ in (\ref{eq: ghat}). This expansion procedure is termed the Dynamic Subspace Expansion (\textbf{DSE}). As the subspace $V$ grows in size, 
\begin{itemize}
    \item the operation $V \otimes V$ becomes exponentially more time intensive (even if previously computed terms are saved),
    \item but $V\!\!\leftarrow\![V \, {\bm x}]$ adds marginally less important basis terms.
\end{itemize}
Accordingly, the first elements of $\delta\hat{{\bf x}}_{c}$ tend to be orders of magnitude larger than the final elements. Since $V \otimes V$ is \textit{only} used in constructing ${\hat H}$, which in-turn is used to compute the quadratic terms in the residual function $\hat{\bf g}$ from (\ref{eq: ghat}), we can curtail the growth of ${\hat H}$ by choosing to only keep the expansions associated with first $(n_q)^2$ quadratic terms of $\delta\hat{{\bf x}}_{c}\otimes\delta\hat{{\bf x}}_{c}$. This can be done without reasonably compromising the quality of (\ref{eq: ghat}), and it is implemented in line 9 of Alg. \ref{alg:DSE}.

{The hyperparameters $\epsilon_B$ and $n_q$ from Alg. \ref{alg:DSE}, which control the expansion size of $V$ and $V \otimes V$, respectively, can be selected via direct experimentation. Alternatively, their respective sizes can be estimated by taking a singular value decomposition of a representative voltage data matrix from the system (this will be done in Fig. \ref{fig:SVD_Analysis}). The value of $\epsilon_B$ can be estimated by determining where on the curve represents an optimal trade-off between accuracy and complexity. The value of $n_q$ can be estimated by considering the \textit{square} of the singular values: the index of the first squared singular value whose quadratic contribution is less significant than its associated computational burden increase can help estimate the value of $n_q$.} The test results section provides further discussion on the selection of these parameters.

\subsection{Combining the NPFS, RMS, and DSE for Accelerated PPF}
\begin{figure}
    \centering
    \includegraphics[width=1\columnwidth]{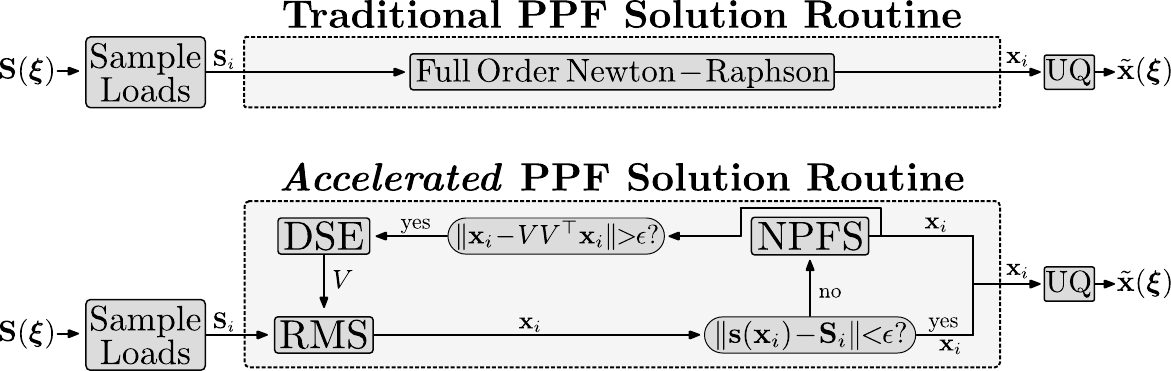}
    \caption{Contrasted are the traditional sampling-based Probabilistic Power Flow (PPF) {solution routine} (top panel), and the Accelerated Probabilistic Power Flow (APPF) {solution routine} (bottom panel). {In both cases, uncertain distribution grid load profiles are first sampled and then passed into their respective solvers. Once a solution is obtained, an external procedure provides UQ.} In the APPF, a reduced modeler first attempts to solve power flow. If it fails, the full-order Neumann series solver is used. If the resulting solution isn't in the basis $V$, it is dynamically added, and the reduced model is updated.}
    \label{fig:PPF}
\end{figure}

We now incorporate the NPFS, the RMS, and the DSE into one coherent routine {known as the Accelerated Probabilistic Power Flow (\textbf{APPF}).} On the surface, this solver behaves like other sampling-based PPF {solvers}: it loops over various sampled load profiles and solves power flow for each one. Such architecture can be seen in the top panel of Fig. \ref{fig:PPF}, where a traditional Newton-Raphson solver (i.e., (\ref{eq: power_flow_it})) is used. Our routine, however, which is shown in the bottom panel of Fig. \ref{fig:PPF}, {proceeds in the following way:
\begin{itemize}
    \item First, the APPF attempts to solve the power flow problem using the RMS.
    \item If the resulting residual from the RMS is too large, then the inadequate solution is passed to the NPFS. The NPFS is capable of solving the power flow problem to $\epsilon_N$ accuracy (on par with standard Newton).
    \item The resulting solution is then passed to the DSE, and the subspace $V$ is potentially updated. Furthermore, the DSE updates the ROM of the power flow problem, and it passes this down to the RMS.
\end{itemize}}
{Further details are given in Algorithm \ref{alg:APPF}, which is termed the APPF. Notably, the ROM is initialized as a 1 dimensional system, so initially, and by design, it is not powerful enough to solve the power flow problem on its own. As the ROM grows in size over many power flow sample iterations, the solutions produced by the RMS become increasingly more accurate. We note that any sufficiently inaccurate solution produced by the RMS is always improved upon by the NPFS, so the APPF is always able to produce accurate power flow solutions. As demonstrated in the following section, the RMS eventually becomes so accurate that NPFS and DSE routines are bypassed entirely. We also note that, in Algorithm \ref{alg:APPF}, the APPF is not told \textit{a priori} how many dimensions the final ROM should include; instead, the ROM dynamically expands whenever the APPF encounters a solution from the NPFS which its current reduced model could not construct.}

Both Algorithm \ref{alg:APPF} and the APPF diagram in Fig. \ref{fig:PPF} require a series of load profile samples ${\bf S}_i$ to be given as inputs to our solver. Our algorithm, though, is agnostic to how these inputs are chosen. If the underlying load distributions in the uncertainty set are uncorrelated, then the distributions can be sampled independently. If the distributions are correlated, however, their associated joint distributions can be used to sample and properly capture this correlation.

\begin{algorithm}
\caption{Dynamic Subspace Expansion (DSE)}\label{alg:DSE}
{\small

\begin{algorithmic}[1]

\Function{$[V,\hat{J}{...}]\!\leftarrow$DSE}{$V\!,\hat{J},\hat{G},\hat{H},{H_V}\!,\delta\hat{\bf x}_c,\hat{\bf s}_0,{\bf s}_0,{\bf x},J_{\!c0},H_{\!c}$}
\State ${\bf x}_c\leftarrow$ Polar-to-Cartesian$\{{\bf x}\}$

\State ${\bm x} \leftarrow {\bf x}_c - V V^{\top} {\bf x}_c$
   
  \If{$\Vert{\bm x}\Vert >$ tolerance $\epsilon_B$}
    
    \State $\delta\hat{\bf x}_c\leftarrow {\bm [}\delta\hat{\bf x}_c^{\top}\; \Vert{\bm x}\Vert{\bm ]}^{\top}$
    
    \State ${\bm x} \leftarrow {\bm x} / \Vert{\bm x}\Vert$
      
    \State ${\bm x}_{\! J}\leftarrow J_{c0}\bm{x}$
    
    \State $\hat{G}\leftarrow\left[\!\!\!\begin{array}{cc}
    \hat{G} & \hat{J}{\bm x}_{\! J}\\
    {\bm x}_{\! J}^{\top}\hat{J} & {\bm x}_{\! J}^{\top}{\bm x}_{\! J}
    \end{array}\!\!\!\right]$
    \If{ ${\rm size} \{\delta \hat{\bf x}_c\} \le n_q $}
    
   \State $V_{k}\leftarrow{\rm sort}\{\bm{x}\otimes V,{\bm{x}}\otimes\bm{x}\}$
   
   \State $\hat{H}\leftarrow{\rm sort}\left[\!\!\!\begin{array}{cc}
    \hat{H} & \hat{J}^{\top}H_{c}V_{k}\\
    \bm{x}_{J}^{\top}{H_V} & \bm{x}_{J}^{\top}H_{c}V_{k}
    \end{array}\!\!\!\right]$
    
    \State {${H}_{V}\leftarrow{\rm sort}{\bm [}{ H}_{V}\;\;{H}_{c}{V}_{k}{\bm ]}$}
    
    \EndIf {\bf end}
    
    \State $\hat{{\bf s}}_0\leftarrow{\bm [}\hat{{\bf s}}_0^{\top}\;{\bm x}_{\! J}^{\top}{{\bf s}}_0{\bm ]}^{\top}$

    \State $\hat{J}\leftarrow{\bm [}\hat{J}\;\;{\bm x}_{\! J}{\bm ]}$
    
    \State $V \leftarrow {\bm [}V\;\; {\bm x}{\bm ]}$
                
  \EndIf {\bf end}

\State \Return ${ V},\hat{ J},{\hat{ G},\hat{ H},{ H}_V,\delta\hat{\bf x}_c,{\hat {\bf s}}_0}$

\EndFunction
  
\end{algorithmic}}
\end{algorithm}

\begin{algorithm}
\caption{Accelerated Probabilistic Power Flow (APPF)}\label{alg:APPF}

{\small \textbf{Require:} Matrix factors $\mathcal L$, $\mathcal U$, $\mathcal P$ from (\ref{eq: LDL}); initial voltage solution ${{\bf x}}_0$ of nominal power injection ${\bf s}_0$; nominal reduced power flow function ${\bf s}(\cdot)$; reduced power flow Hessian $H_c$ and Jacobian $J_{c0}$ evaluated at ${\bf x}_0$, specified power injection profiles ${\bf S}_i$ for each $i=1,2,...,M$

\textbf{Ensure:} Each solution $ {\bf x}_i$ satisfies ${{\bf s}}({\bf x}_i)  \approx {{\bf S}}_i$

\begin{algorithmic}[1]

\Function{$[{\bf x}_1,{\bf x}_2,...,{\bf x}_M] \leftarrow\,$APPF }{${\mathcal L},{\mathcal U},{\mathcal P},{{\bf S}},{\bf x}_0,{\bf s}_0,J_{c0},H_c$}

\State ${\bf x}_{c0} \leftarrow $ Polar-to-Cartesian$\{{\bf x}_{0}\}$

\State ${\hat {\bf x}}_{c0} \leftarrow \Vert {\bf x}_{c0} \Vert$

\State $\delta{\hat {\bf x}}_{c}\leftarrow0$

\State $V\leftarrow {\bf x}_{c0}/\Vert {\bf x}_{c0} \Vert$

\State ${\hat J} \leftarrow J_{c0} V$

\State $\hat{\bf s}_0 \leftarrow {\hat J}^{\top}{\bf s}_0$

\State ${\hat G}\leftarrow {\hat J}^\top {\hat J}$

\State {${H}_{V}\leftarrow H_{c}(V\otimes V)$}

\State ${\hat H} \leftarrow \hat{J}^{\top}{{H}_{V}}$

\State $i\leftarrow 1$

\While{$i \le M$}

\State ${\hat {\bf S}}_i \leftarrow {\hat J}^\top {\bf S}_i $

\State $[\delta\hat {\bf x}_c,{\bf x}_i]\leftarrow\,$RMS($V,{\hat{\bf s}}_0,{\hat G},{\hat H},\hat{\bf x}_{c0},\delta\hat{\bf x}_{c},{\hat {\bf S}}_i$)

\If{$\Vert {\bf s}({\bf x}_i) - {\bf S}_i \Vert_{\infty}>$ tolerance $\epsilon_N$}

\State ${\bf x}_i \leftarrow$NPFS(${\mathcal L},{\mathcal U},{\mathcal P},{{\bf S}}_i,{\bf x}_i$)

\State $[V,\hat{J},\hat{G},\hat{H},{H_V},\delta\hat{\bf x}_c,{\hat {\bf s}}_0]\leftarrow$DSE$(V,\!\hat{J},\!\hat{G},\!\hat{H},{H_V}...)$

\EndIf {\bf end}

\State $i \leftarrow i + 1$

\EndWhile {\bf end}

\State \Return ${\bf x}_1,{\bf x}_2,...,{\bf x}_M$

\EndFunction

\end{algorithmic}}
\end{algorithm}

\section{Test Results}\label{Sec:Results}
\begin{figure}
    \centering
    \includegraphics[bb=30bp 30bp 492bp 210bp,clip,width=0.95\columnwidth]{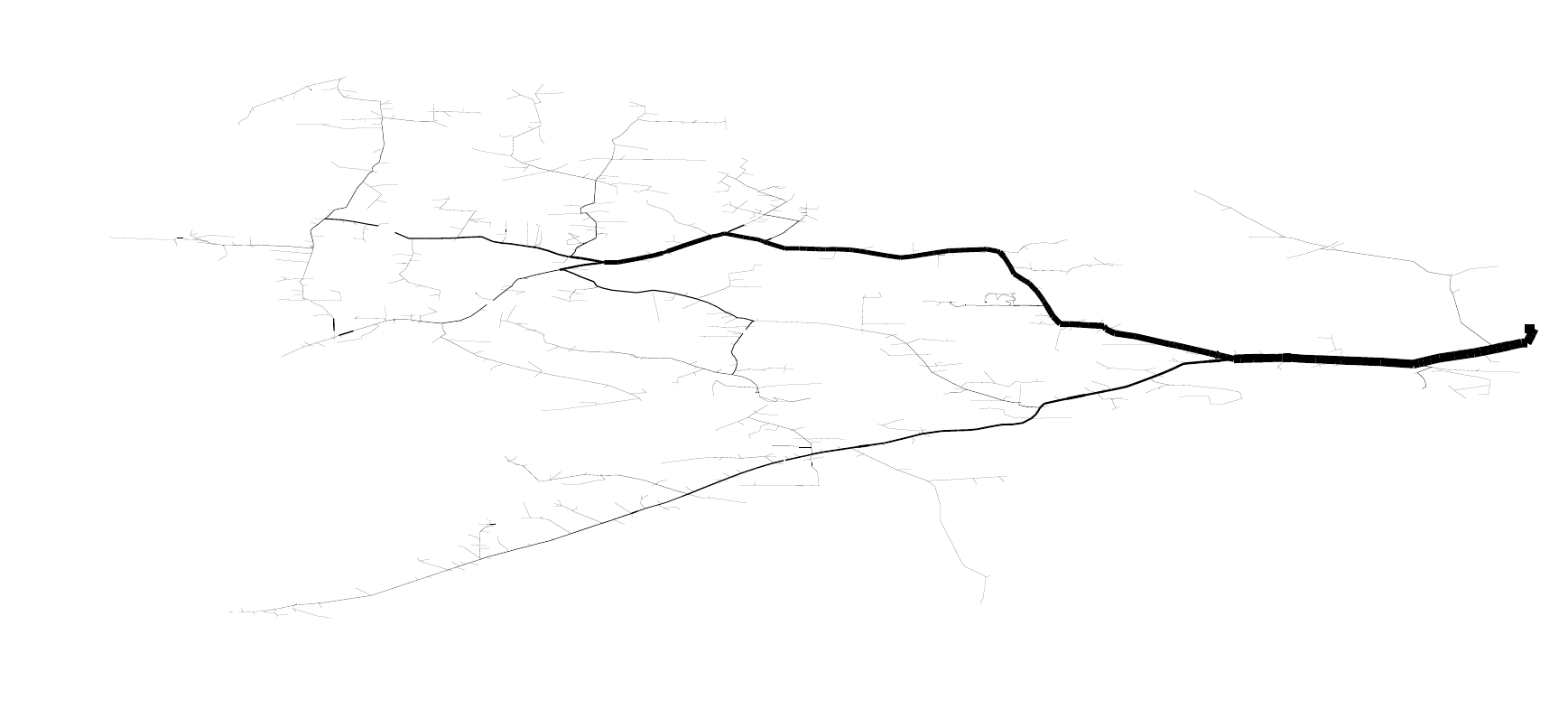}
    \caption{{Circuit diagram of the 8500-node distribution circuit. In this network, there is a single point where positive power is injected (the substation).}}
    \label{fig:Map_Blank}
\end{figure}

In this section, we present test results which were collected on the {unbalanced} IEEE 8500-node test feeder{\cite{Arritt:2010}}. {The circuit diagram associated with this distribution circuit is shown in Fig. \ref{fig:Map_Blank}.} To construct this test case, we used OpenDSS~\cite{Dugan:2011} to carefully export the admittance matrix, nominal loading values, transformer tap ratios, transformer configurations ($\Delta\!:\!{\rm Y}$, ${\rm Y}\!:\!{\rm Y}$, split-phase) and base voltage levels of this network to MATLAB. We then per-unitized the network with a base power of 100kW and base voltages of 66.5kV, 7.2kV and 120V on the appropriate buses. In our tests, all tap ratios and switch configurations were assumed fixed. Additionally, the nominal 3-phase {substation} voltage was assumed static across all trials. With this per-unitization, the largest nominal load current was 0.51 pu, meaning (\ref{eq: ratio_proof}) was satisfied by many orders of magnitude. 

In the following subsections, we present PPF test results collected under two opposing assumptions. In the first scenario, we assume the load sampling distributions are fully uncorrelated (i.e., 0\% correlated); in the second scenario, we assume the sampling distributions are fully correlated (i.e., 100\% correlated). By presenting results associated with these two extreme assumptions, we show that our APPF method is capable of handling any arbitrary degree of correlation among the uncertain loads.

\subsection{{Uncorrelated Load Distributions}}
In testing this network, we chose a subset $\mathcal S$, $|\mathcal S|\!=\!25$, of the largest loads in the 8500-node network and assumed an extremely high degree of input variability. Accordingly, we sampled from i.i.d. Gaussian distributions, such that the sampled active and reactive power were generated via ${\{P,Q\}}_{i}^{(s)} ={\{P,Q\}}_{i}(1+\mathcal{N}(\sigma,0)),\;i\!\in\!\mathcal{S}$.
In each case, the loads in $\mathcal S$ were assumed to have distributed energy resource capabilities (i.e., rooftop PV, battery charging/discharging capabilities, etc.). Accordingly, the standard deviation in was set to $\sigma = 1$, meaning the loads could potentially switch sign and become sources in some sampling instances. Since samples were drawn randomly, this sampling routine is called Simple Random Sampling (SRS)~\cite{Hajian:2013}. If desirable, alternative distributions could be used to characterize the uncertainty set, and correlation between the loads can be introduced as well. The assumption of 0\% correlation is in fact a ``worst-case" assumption, because it leads to the system exploring a much larger operational space. This ultimately expands the size of the ROM\footnote{{When the system was tested with an assumption of 100\% correlation among the loads in the uncertainty set, the resulting ROM dynamically expanded to a dimensionality of just $n=6$.}} and slows down its performance.

The sampling procedure was performed 1000 times for each load, the largest of which vary between -110kW and +60kW of active power, for example. The remainder of the loads were left fixed to one half their nominal values. Next, we documented the speed up of the APPF relative to the traditional PPF solver (see top panel of Fig.~\ref{fig:PPF}) in the context of SRS. All simulations were performed using MATLAB R2017b on a Dell XPS laptop, equipped with an Intel i5 CPU @ 2.30GHz and 8 GB of RAM.

\subsubsection{Traditional PPF}
\begin{figure}
    \centering
    \includegraphics[width=1\columnwidth]{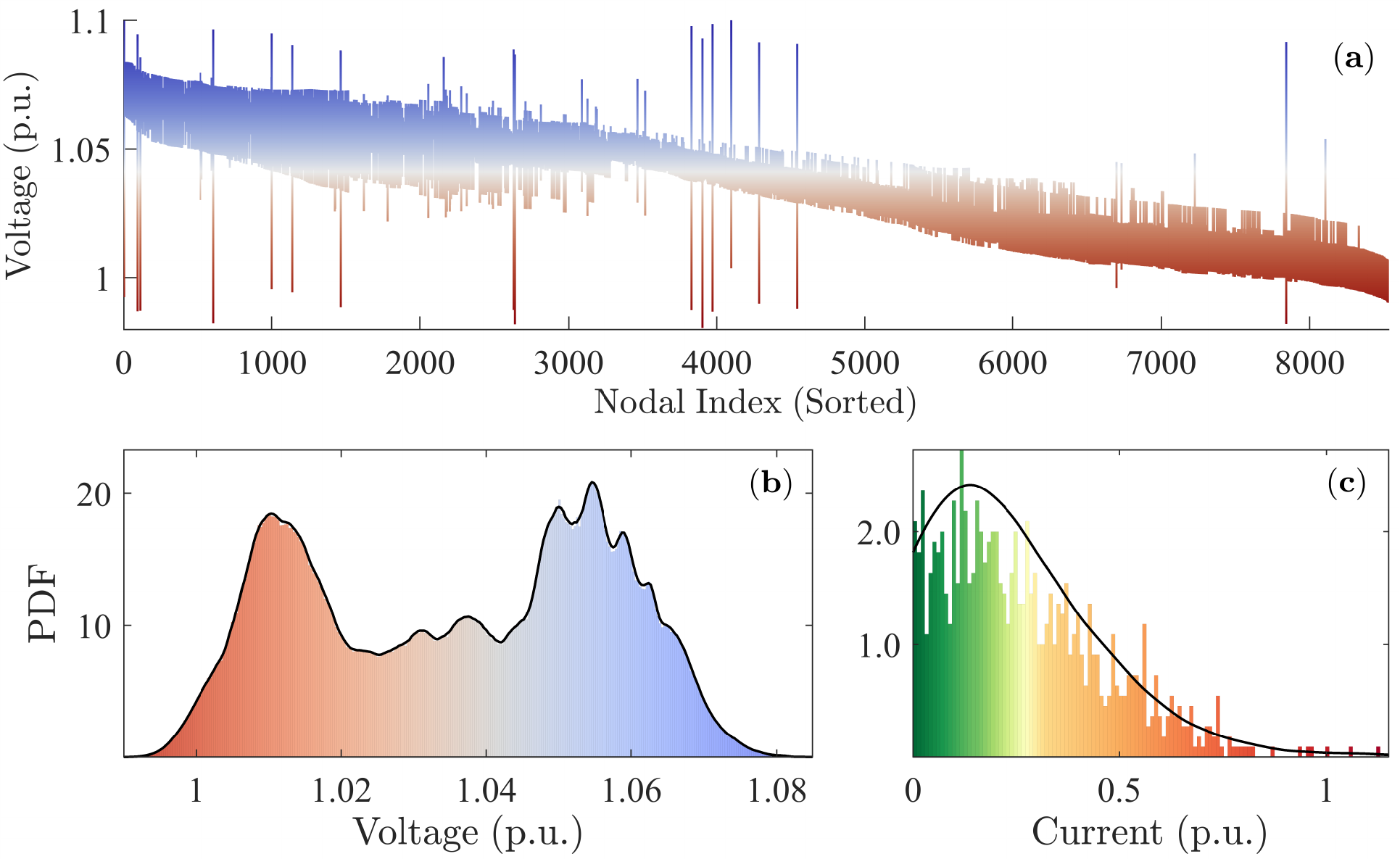}
    \caption{PPF results from the 8500-node network. Panel ($\bf a$) shows the range of output voltages across 1000 trials for each node (sorted for clarity.) Panel ($\bf b$) shows a histogram of all $1000\times8531$ voltage magnitude points across all nodes, with an approximated black PDF curve plotted on top. Panel ($\bf c$) shows a histogram of the current magnitudes flowing on the line connecting nodes 5724 and 8410; an approximated black PDF curve is plotted on top.}
    \label{fig:Voltage_Distribution}
\end{figure}

First, we applied the traditional PPF solver from Fig. \ref{fig:PPF}. The Newton stopping criteria on the power injection residual was set to $\epsilon_N=10^{-4}$, i.e., $\Vert {\bf s}({\bf x}_i) - {\bf S}_i \Vert_{\infty}<10^{-4}$. Physically, this corresponds to conservation of power being satisfied below 10 Watts at each node in the network. Looping over the 1000 loading configurations, Newton required between 2 and 3 steps to converge for all load configurations except for the first one, as shown in panel (${\bf a}$) of Fig. \ref{fig:Iteration_Tracker}. Ultimately, the full simulation required $\sim$190 seconds to run, meaning each power flow lasted about 0.19 seconds. Sample results from the simulation are shown in Fig. \ref{fig:Voltage_Distribution}. These plots show voltage and current distributions in the network, and they represent some of the ways that characterizing the probabilistic output from a PPF routine can be useful. The results generated by the traditional PPF routine, for all practical purposes, are \textit{identical} to the results generated by the APPF routine in the following subsection, i.e., the data shown in Fig. \ref{fig:Voltage_Distribution} could be generated by either process. {Ultimately, this is because both power flow algorithms are held to the same convergence tolerance, $\epsilon_N$ (for the APPF routine, see line 15 of Alg. \ref{alg:APPF}, followed, if necessary, by line 4 of Alg. \ref{alg:NPFS}).}

\subsubsection{APPF}
In order to run the APPF, three additional tolerances were needed: the RMS convergence tolerance $\hat{\epsilon}_{N}$ from Alg. \ref{alg:RMS}, which was set to $\hat{\epsilon}_N=10^{-5}$, the expansion curtailment constant $n_q$ from Alg. \ref{alg:DSE}, which was set to $n_q=37$, and the basis expansion tolerance $\epsilon_B$ from Alg. \ref{alg:DSE}. The choice of $\epsilon_B$ was particularly important: if set too large, the basis would never fill up and the RMS would perform poorly, but if set too small, the basis would fill up endlessly and slow the RMS down considerably. In testing the 8500-node network, we found $\epsilon_B=10^{-4}$ to be an effective compromise. {Selecting the proper value of this parameter can generally be achieved after brief experimentation with the distribution system in question.}

\begin{figure}
    \centering
    \includegraphics[width=1\columnwidth]{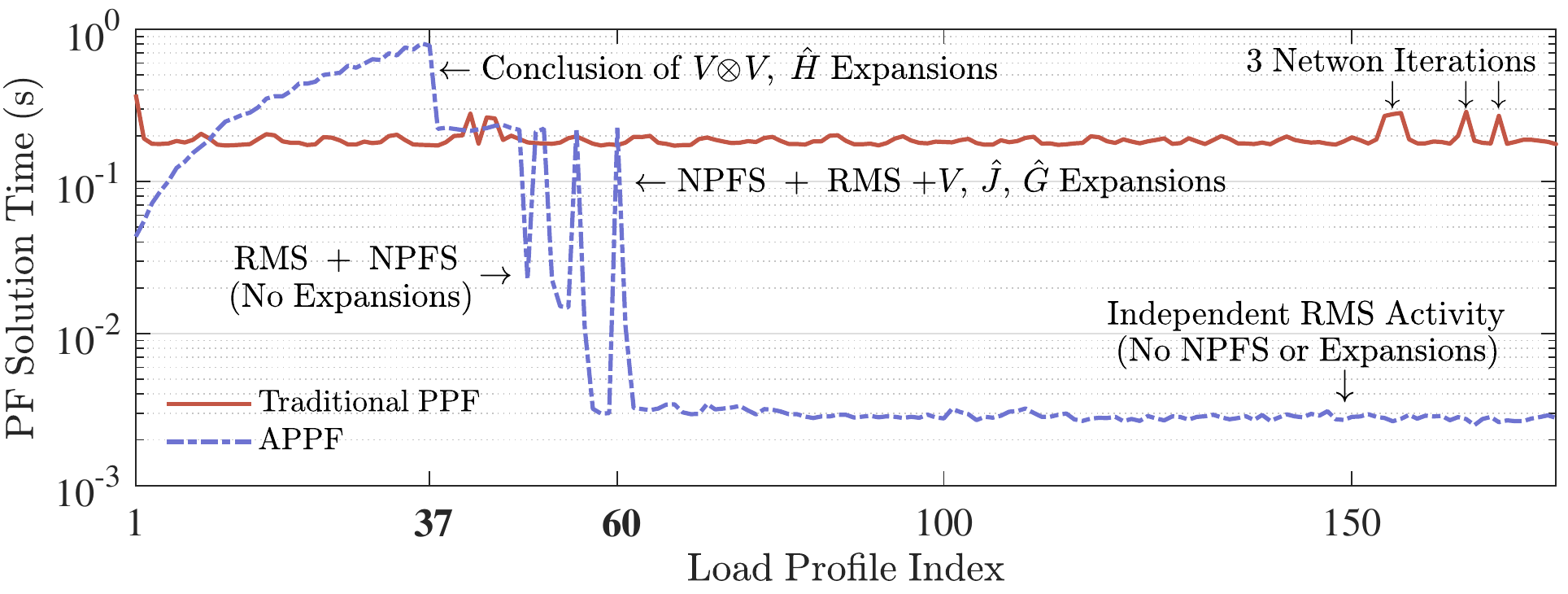}
    \caption{Shown is timing analysis for the traditional Newton-based PPF solver versus the APPF solver over the first 175 load profiles. Notably, the traditional PPF solver has a fairly constant ``solve speed", usually requiring two Newton iterations (although the locations of three iterations are marked in the upper right). At load profile $n_q=37$, the APPF stopped building $H_c$, which was becoming very time intensive. Between load profiles 38 and 60, the NPFS and RMS worked together to continue building out the basis $V$ and solving the system. After load profile 60, the RMS quickly solved the system on its own without the NPFS, and no more basis expansion was needed.}
    \label{fig:TimingAnalysis}
\end{figure}

With these tolerances set, the APPF simulation required $\sim$20 seconds to run. {Ultimately, the APPF was held to (and satisfied) the same convergence criteria as the traditional PPF: conservation of power was satisfied at each node by a margin of less than $\epsilon_N=10^{-4}$.} Relative to the traditional PPF, though, the APPF ran almost \textit{10x faster}. The super-majority of this time, though, was spent construing matrix ${\hat H}=\hat{J}^{\top}H_{c}(V\!\otimes\!V)$, which occurred as subspace $V$ was being intermittently constructed during the first 60 or so load profile iterations. Once $V$ was sufficiently filled out, the RMS could solve the power flow problem without any help from the NPFS (and without any more basis expansions). The final 940 load profiles were solved in 2.90 seconds. Relative to the final 940 solves of the traditional PPF, the APPF ran \textit{$\sim$61x faster}. The full timing breakdown is graphically portrayed in Fig. \ref{fig:TimingAnalysis}. {The speed up from the traditional PPF to the APPF is only related to speeding up the middle box in Fig.~\ref{fig:TPPF_Routine}; that is, the power flow problem is being solved faster. There are many advanced sampling and advanced UQ tools (see Introduction) which could speed up the PPF process even more, but ultimately, these methods are constrained by how fast the power flow problem can be solved. Our methods, therefore, show great promise in speeding up any sampling-based PPF routine.}

\begin{figure}
    \centering
    \includegraphics[width=1\columnwidth]{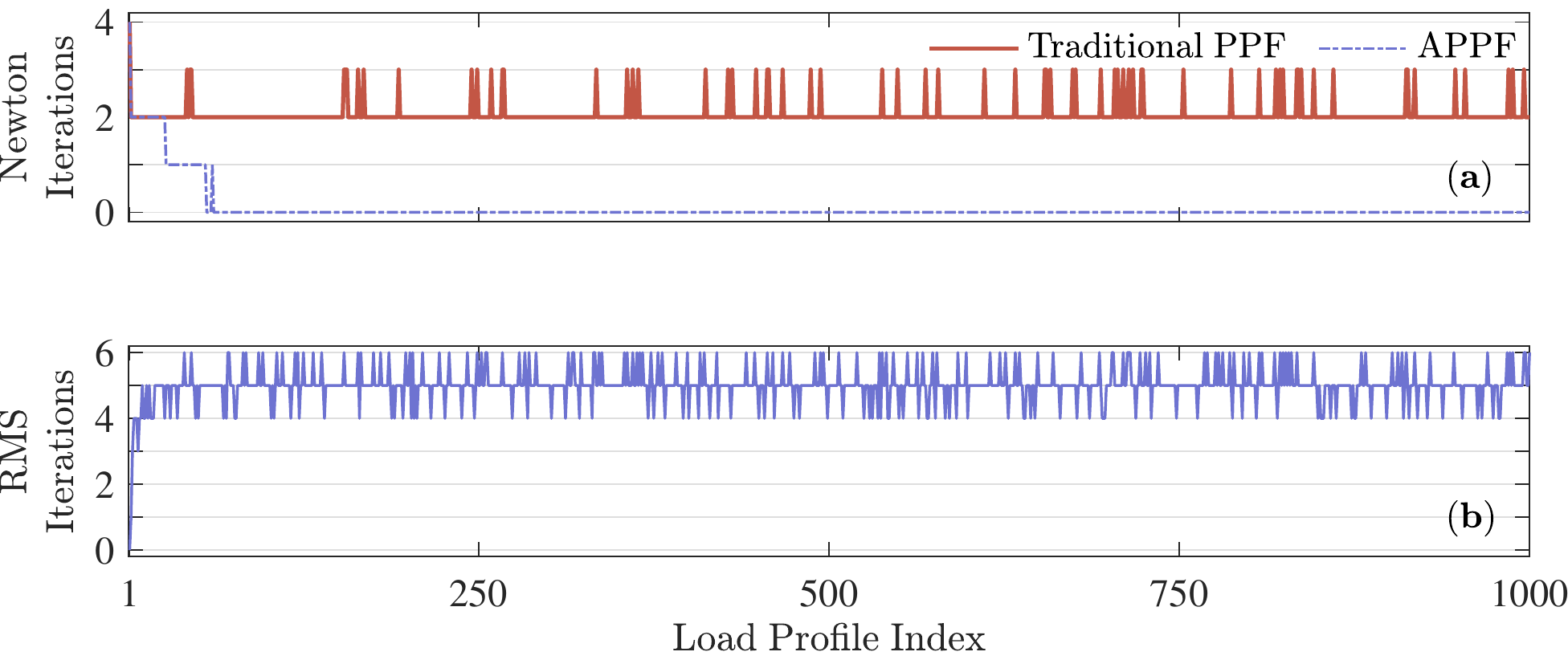}
    \caption{Iteration tracker for the experiment with fully \textit{uncorrelated} loads. Panel ($\bf a$) shows the number of Newton iterations performed by traditional PPF for each new load profile vs. the number of Newton iterations taken by the NPFS in the APPF. Panel ($\bf b$) shows the number of iterations taken by the RMS inside the APPF.}
    \label{fig:Iteration_Tracker}
\end{figure}

It is also instructive to consider how many Newton iterations the traditional PPF and the APPF solvers were required to perform on the full nonlinear system. This is shown in panel ($\bf a$) of Fig. \ref{fig:Iteration_Tracker}. As the APPF runs and $V$ fills up, the number of required Newton iterations by the NPFS in Alg. \ref{alg:NPFS} drops from 4, to 3, to 2, to 1, to 0. When the APPF does take a Newton step, though, it is much faster than the traditional PPF Newton step, due to the Neumann expansion. On average:
\begin{itemize}
    \item Traditional PPF Newton step time (mean): \textbf{0.11 seconds}
    \item APPF Newton step time via NPFS (mean): \textbf{0.02 seconds}
\end{itemize}
\indent Panel ($\bf b$) of Fig. \ref{fig:Iteration_Tracker} shows the number of iterations, usually 5 or 6, taken by the RMS as the load profiles are processed. This relatively large number of iterations is due to the recycling of approximate reduced Jacobian $\hat G$. If $\hat G$ was exactly computed at each step, fewer iterations would be necessary. Such updating, though, is far more expensive than adding additional iterations, so we tolerate the high iteration count in Fig. \ref{fig:Iteration_Tracker}.

\begin{figure}
    \centering
    \includegraphics[width=1\columnwidth]{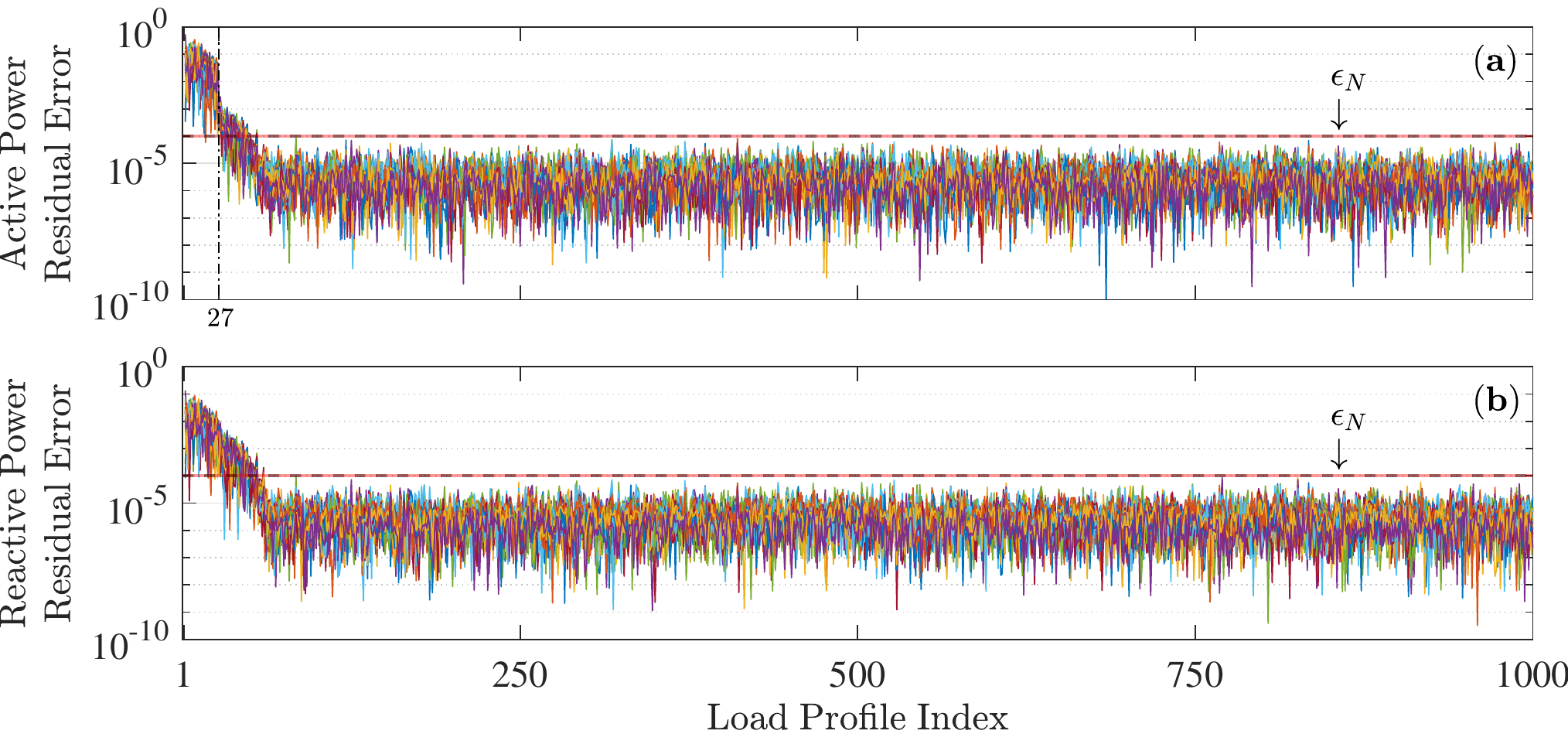}
    \caption{Residual active and reactive power (panels ($\bf a$) and ($\bf b$), respectively) at the 25 perturbed load buses in the 8500-node network \textit{after} the RMS has converged. As $V$ expands, the RMS is able to consistently drive the residual at each of these buses below the stopping criteria $\epsilon_N$.}
    \label{fig:Residual_Powers}
\end{figure}

\begin{figure}
    \centering
    \includegraphics[width=1\columnwidth]{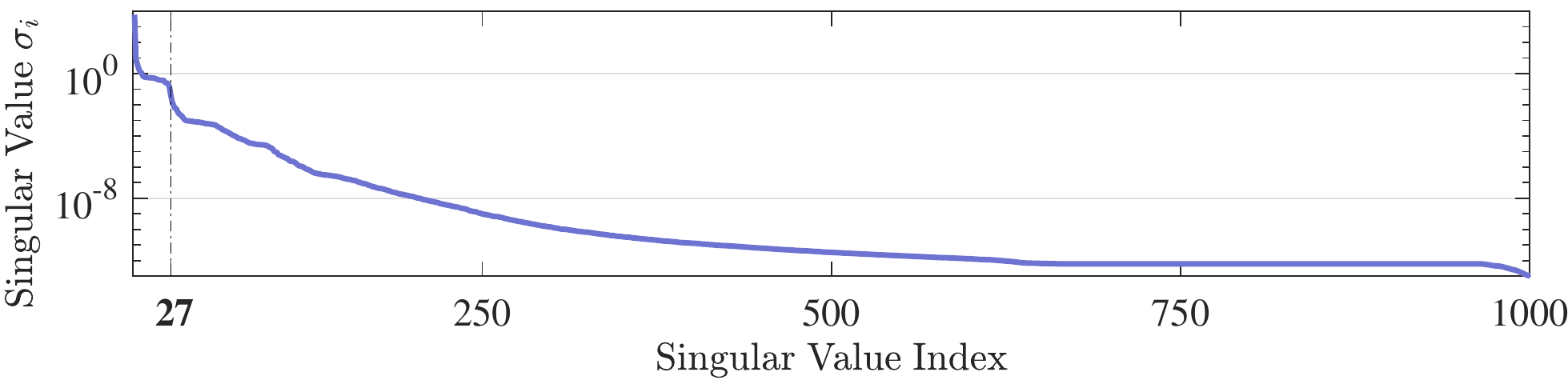}
    \caption{{Plot of the singular values $\sigma$ of matrix $W=[{\bf x}_{1},\,{\bf x}_{2},...,{\bf x}_{1000}]$.}}
    \label{fig:SVD_Analysis}
\end{figure}

As the APPF solver ran, the basis $V$ dynamically expanded to include 52 orthonormal columns, giving it an ultimate dimension of $(2\cdot8531)\times 52$. As $V$ expanded, the RMS became increasingly effective at solving the power flow problem without any help from the NPFS. This is shown very clearly by Fig. \ref{fig:Residual_Powers}, which shows how the RMS output residual decreases as the solver cycles through the load profiles. It is interesting to note the salient ``residual cliff" in panel ($\bf a$), quite clearly located at load profile 27. To further explore its significance, we stacked the voltage solution vectors ${\bf x}_i$ (found by traditional PPF) inside data matrix $W=[{\bf x}_{1},\,{\bf x}_{2},...,{\bf x}_{1000}]$. We then took the SVD of $W$, i.e., ${\bm \sigma}={\rm svd}(W)$. The results are shown in Fig. \ref{fig:SVD_Analysis}, which clearly shows a steep drop-off after the $27^{\rm th}$ largest singular value. This provides a nice qualitative explanation for the residual cliff in Fig. \ref{fig:Residual_Powers}: the first 27 columns of subspace $V$ capture the most important features of the voltage profile, where ``important" is quantified by the magnitude of a corresponding singular value.

\subsection{{Fully Correlated Load Distributions}}
{In contrast to the previous experiment, where loads were fully uncorrelated, we now enforce the uncertain load distributions to be 100\% correlated. That is, each load in the uncertainty set $\mathcal S$ was perturbed by a percentage sampled from the same distribution; this is equivalent to scaling these loads by a common (but random) factor $\alpha$.
While this does represent an unusually high degree of correlation, we use this example to showcase how the APPF performs when high degrees of correlation are present.}

\begin{figure}
    \centering
    \includegraphics[width=1\columnwidth]{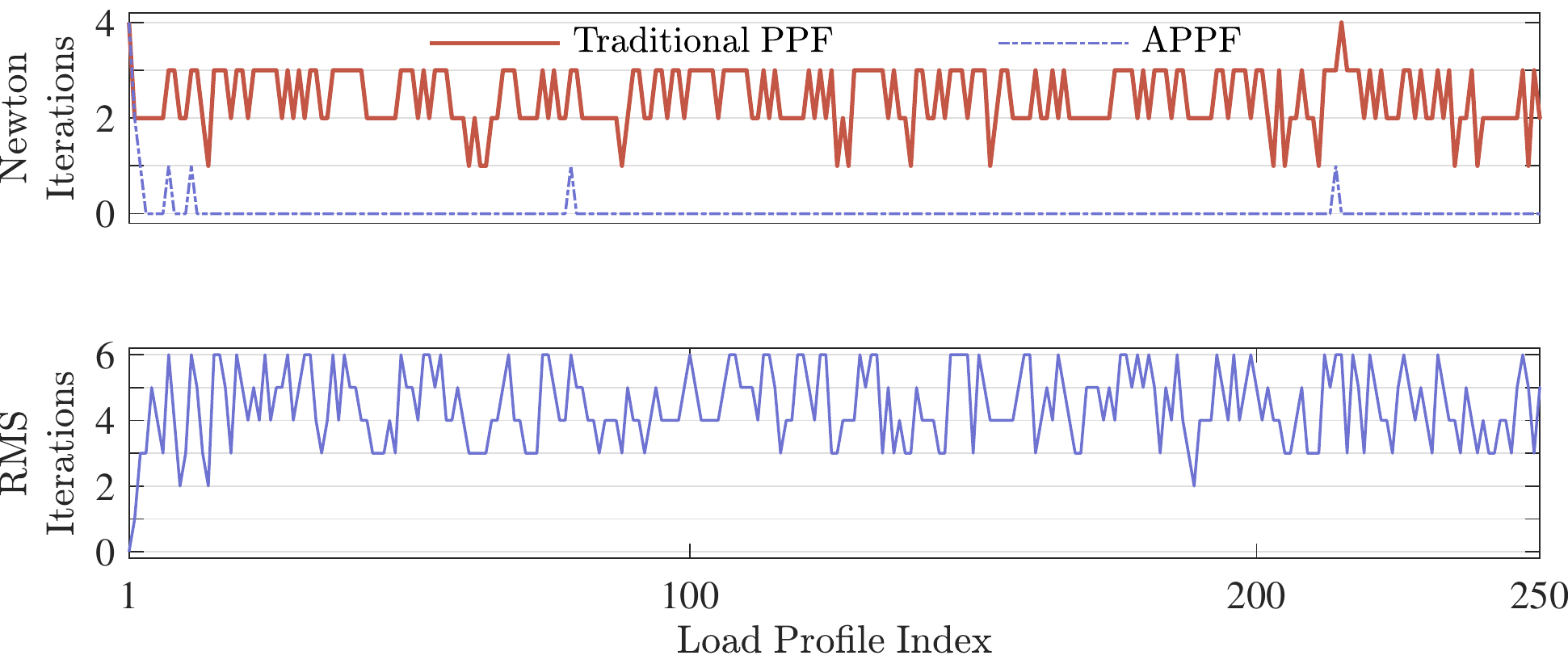}
    \caption{{Iteration tracker for the experiment with fully \textit{correlated} loads. Panel ($\bf a$) shows the number of Newton iterations performed by traditional PPF for each new load profile vs. the number of Newton iterations taken by the NPFS in the APPF. Panel ($\bf b$) shows the number of iterations taken by the RMS inside the APPF.}}
    \label{fig:Iteration_Tracker_100p}
\end{figure}

\begin{figure}
    \centering
    \includegraphics[width=1\columnwidth]{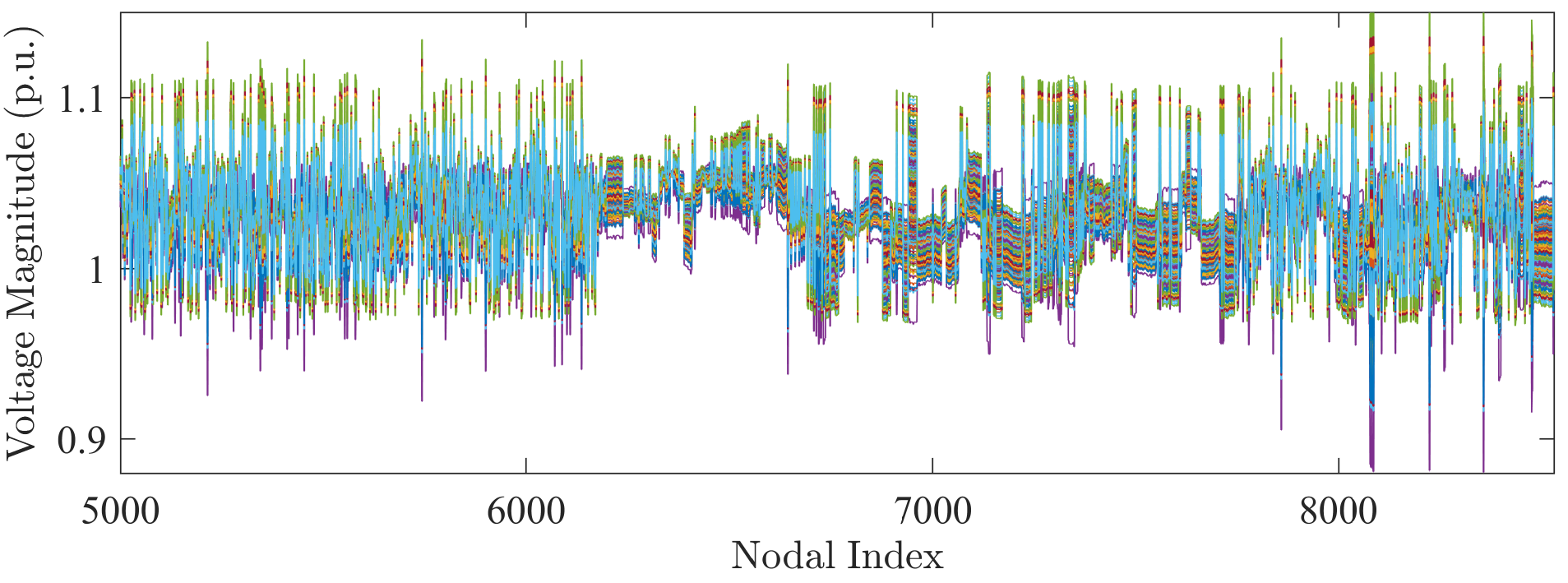}
    \caption{Nodal voltage magnitude solutions associated with the fully correlated load experiment. Each trace represents one of the 1000 power flow solutions; voltage magnitudes for only the final 3531 nodes are shown.}
    \label{fig:Voltage_Magnitudes_100p}
\end{figure}

{In this case, the traditional PPF solver solved the associated 1000 power flow problems in 101.44 seconds; sample voltage magnitude results are shown in Fig. \ref{fig:Voltage_Magnitudes_100p}. The APPF, however, solved the same power flow problems in 0.607 seconds. Remarkably, this represents a computational speed up of $\sim$167.2x. This massive computational speed up is primarily due to the fact that the ROM only grew into a 6-dimensional system. That is, the reduced state vector $\hat {\bf x}$  only had 6 dimensions. Thus, the RMS could solve the associated reduced system very rapidly. In Fig. \ref{fig:Iteration_Tracker_100p}, we demonstrate why the ROM was able to grow into a fully expressive model so quickly. The top panel shows that APPF Newton iterations (as taken by the Neumann solver) were only necessary at load profiles 1, 2, 3, 8, 12, 79, and 214. At all other load profiles, the RMS was able to construct a fully accurate solution on its own, and no basis expansions were necessary. It is notable that a 6-dimensional ROM system was able to fully capture the nontrivial voltage profile demonstrated in Fig. \ref{fig:Voltage_Magnitudes_100p}.}

\section{Conclusion}\label{Sec:Conclusion}
In this paper, we showed how to speed up ``any'' sampling-based PPF solver by 1) leveraging the low-rank nature of distribution network voltage profiles to generate a surrogate ROM, and 2) exploiting the ``small" nature of distribution network loads and applying a custom Neumann series-based method. Our resulting APPF algorithm, which combines both contributions {and is fully invariant to the underlying network topology}, was tested on the 8500-node network, speeding up full-order Newton based PPF methods by up to 10 times. Once the overhead associated with constructing the ROM was cleared, the ROM ran over 60 times faster than full-order Newton {in the experiment with fully uncorrelated loads.}

{Based on the successful test results, we conclude that the APPF could be used in conjunction with either advanced sampling approaches (e.g., adaptive importance sampling or Latin hypercube sampling), or advanced UQ tools (e.g., polynomial chaos or stochastic testing) to further speed up PPF computations. Once deployed, distribution system operators (DSOs) would be the primary beneficiary of the APPF, since it will allow them to solve PPF problems in real time. This will help them to make more informed operation and control decisions related asset utilization, charging schedules, or DER/load control.}

{The test results also highlight important pros and cons associated with the APPF methodology. While it does offer significant acceleration, it can suffer from the curse of dimensionality. Thus, if the uncertainty set is too large, and the system explores too much of the operational space, the ROM will probably grow so large that its computational benefits disappear. Additionally, the method incurs significant overhead when building up the ROM, as depicted by the blue curve in Fig. \ref{fig:TimingAnalysis}. Thus, for situations where only a moderate number of system solutions are needed (e.g., order 10-100), APPF will not offer much acceleration (in fact, its overhead may make it slower than conventional methods{)}. In situations where 100s or 1000s of power flow solutions are needed, though, the APPF will add significant value.}

We note that the Neumann-based power flow solver is a novel contribution in its own right, and it can be used to solve power flow independently of the ROM. As shown in the test results section, Neumann can speed up the power flow problem by up to a factor of 5. And as previously described, it can be easily combined with other power flow methods (e.g., fast decoupled power flow) to achieve even faster speeds. The NPFS can thus offer computational benefits to many problems which utilize distribution power flow solutions, e.g., future expansion planning, network simulation, and even optimal power flow methods.

Future work will extend these methods to the distribution system state estimation (DSSE) problem, and it will explore the applicability of these methods to transmission network problems. The Neumann-based power flow solver, which exploits small load currents and a single slack-bus injection point, was specifically constructed to solve the distribution grid power flow problem; thus, its applicability to the transmission power flow problem must be investigated and cannot be guaranteed. The proposed projection based MOR approach, however, makes no such assumptions and therefore could be directly extended to the transmission power flow problem. The meshed nature of transmission grids may generally lead to a higher dimensional ROM, so further testing will be needed in order to clarify its effectiveness in this context.

{
\appendices

\section{}\label{AppA}
\begin{proof}
For $(\mathcal{D}_{c}+\mathcal{L}_{c}\mathcal{U}_{c})^{-1}$ to be approximated by a Neumann series, then $\rho((\mathcal{L}_{c}\mathcal{U}_{c})^{-1}\mathcal{D}_{c})<1$ must hold. Expanding,
\begin{subequations}
\begin{align}
\!\!\! \rho((\mathcal{L}_{c}\mathcal{U}_{c})^{-1}\mathcal{D}_{c}) & =\rho(({E}^{\top}Y_{l}{E})^{-1}{\mathtt d}(\tilde{{\bf V}})^{-1}{\mathtt d}(\tilde{{\bf I}}^{*}))\\
 & \le \rho(({E}^{\top}Y_{l}{E})^{-1})\rho({\mathtt d}(\tilde{{\bf I}}^{*}))\\
 & =\frac{{\rm max}\{|\tilde{{\bf I}}|\}}{\rho({E}^{\top}Y_{l}{E})},\label{eq: ratio_proof}
\end{align}
\end{subequations}
where $\rho({\mathtt d}(\tilde{{\bf V}})^{-1}) \approx 1$ is assumed due to per-unitization.
\end{proof}

\section{}\label{AppB}
The expansion of the Cartesian coordinate power flow equations has a Jacobian which is equal to (\ref{eq: RPFJ}), but with the elimination of the polar-to-Cartesian conversion matrix $R(\cdot)$:
\begin{align}\label{eq: RPFJ_c}
{J}_{c}=(\langle{\mathtt d}({{\bf I}}_{\rm r}-j{{\bf I}}_{\rm i})\rangle+\langle{\mathtt d}({{\bf V}}_{\rm r}+j{{\bf V}}_{\rm i})\rangle N\langle{Y}_{b}\rangle).
\end{align}
Notably, (\ref{eq: RPFJ_c}) is a \textit{linear} function of Cartesian voltage coordinates, to the Hessian ${H}_c\in{\mathbb R}^{n\times n^2}$ is constant. With ${i}^{\rm th}$ unit vector ${\bf e}_i$,
\begin{align}{H}_{c} & =\left[\!\begin{array}{cccccc}
\frac{d{J}_c}{d{{\bf V}}_{{\rm r},1}} & \cdots & \frac{d{J}_c}{d{{\bf V}}_{{\rm r},n}} & \frac{d{J}_c}{d{{\bf V}}_{{\rm i},1}} & \cdots & \frac{d{J}_c}{d{{\bf V}}_{{\rm i},n}}\end{array}\!\right]\\
\tfrac{d{J}_c}{d{{\bf V}}_{{\rm r},i}} & =\langle\mathtt{d}({Y}_{b}^{*}{\bf e}_{i})\rangle\!+\!\langle\mathtt{d}({\bf e}_{i})\rangle N\langle{Y}_{b}\rangle\\
\tfrac{d{J}_c}{d{{\bf V}}_{{\rm i},i}} & =\langle\mathtt{d}(-j{Y}_{b}^{*}{\bf e}_{i})\rangle\!+\!\langle\mathtt{d}(j{\bf e}_{i})\rangle N\langle{Y}_{b}\rangle.
\end{align}

\section{}\label{App_min_r}
The Newton-like algorithm for minimizing $\Vert{\bf g}(V\delta\hat{{\bf x}}_{c})\Vert_2^2$ can be derived by keeping the constant + linear terms of the expansion (\ref{eq: Kron_sep}) and then solving for iterative values of $\delta\hat{{\bf x}}_c$ via Moore-Penrose:
\begin{align}\label{eq: xc_its}
\delta\hat{{\bf x}}_{c}^{(i+1)}=\delta\hat{{\bf x}}_{c}^{(i)}-[(J_{c0}V)^{\top}J_{c0}V]^{-1}(J_{c}V)^{\top}{\bf g}(V\delta\hat{{\bf x}}_{c}^{(i)}).
\end{align}
In solving (\ref{eq: xc_its}), we notice that whenever ${\bf g}(V\delta\hat{{\bf x}}_{c}^{(i)})$ is evaluated, it is left multiplied by $(J_{c0}V)^{\top}$. We therefore define ${\hat J} = J_{c0}V$ and then multiply (\ref{eq: Kron_sep}) through by ${\hat J}^{\top}$:
\begin{align}\label{eq: ghat}
\hat{{\bf g}}(\delta\hat{{\bf x}}_{c})=\hat{{\bf s}}_{0}+\hat{G}\delta\hat{{\bf x}}_{c}+\tfrac{1}{2}\hat{H}(\delta\hat{{\bf x}}_{c}\otimes\delta\hat{{\bf x}}_{c})-\hat{{\bf S}}
\end{align}
where $\hat{{\bf g}}=\hat{J}^{\top}{\bf g}$, $\hat{{\bf s}}_{0}=\hat{J}^{\top}{\bf s}_{0}$, ${\hat G}=\hat{J}^{\top}\hat{J}$, ${\hat H}=\hat{J}^{\top}H_{c}(V\!\otimes\!V)$, and $\hat{{\bf S}}=\hat{J}^{\top}{\bf S}$. Importantly, (\ref{eq: ghat}) is a \textit{square} system, i.e.,  it has $q$ equations and $q$ variables in $\delta\hat{{\bf x}}_{c}$. {Furthermore, this reduced system is also nonlinear, meaning the nonlinearity of the power flow equations was not lost in the projection.} Applying a Newton-like method to (\ref{eq: ghat}) yields the iterative routine (\ref{eq: step_xc_Newton}). Notably, $\hat J$, $\hat G$, and $\hat H$ are \textit{constant} matrices and do not need to be updated at each step $\delta{\hat{\bf x}}_{c}^{(i)}$.

\begin{remark}
Because (\ref{eq: ghat}) is a determined system, the residual ${\hat{\bf g}}$ can be driven to ${\bf 0}$. This residual, though, is merely a \textit{projection} of the true residual ${{\bf g}}$ of (\ref{eq: resid_func}) into the low-rank space $(J_{c}V)^{\top}$. Therefore, $\hat{{\bf g}}={\bf 0}$ does not imply ${{\bf g}}={\bf 0}$.
\end{remark}
}

\bibliographystyle{IEEEtran}
\bibliography{PPF}

\begin{IEEEbiography}[{\includegraphics[width=1in,height=1.25in,clip,keepaspectratio]{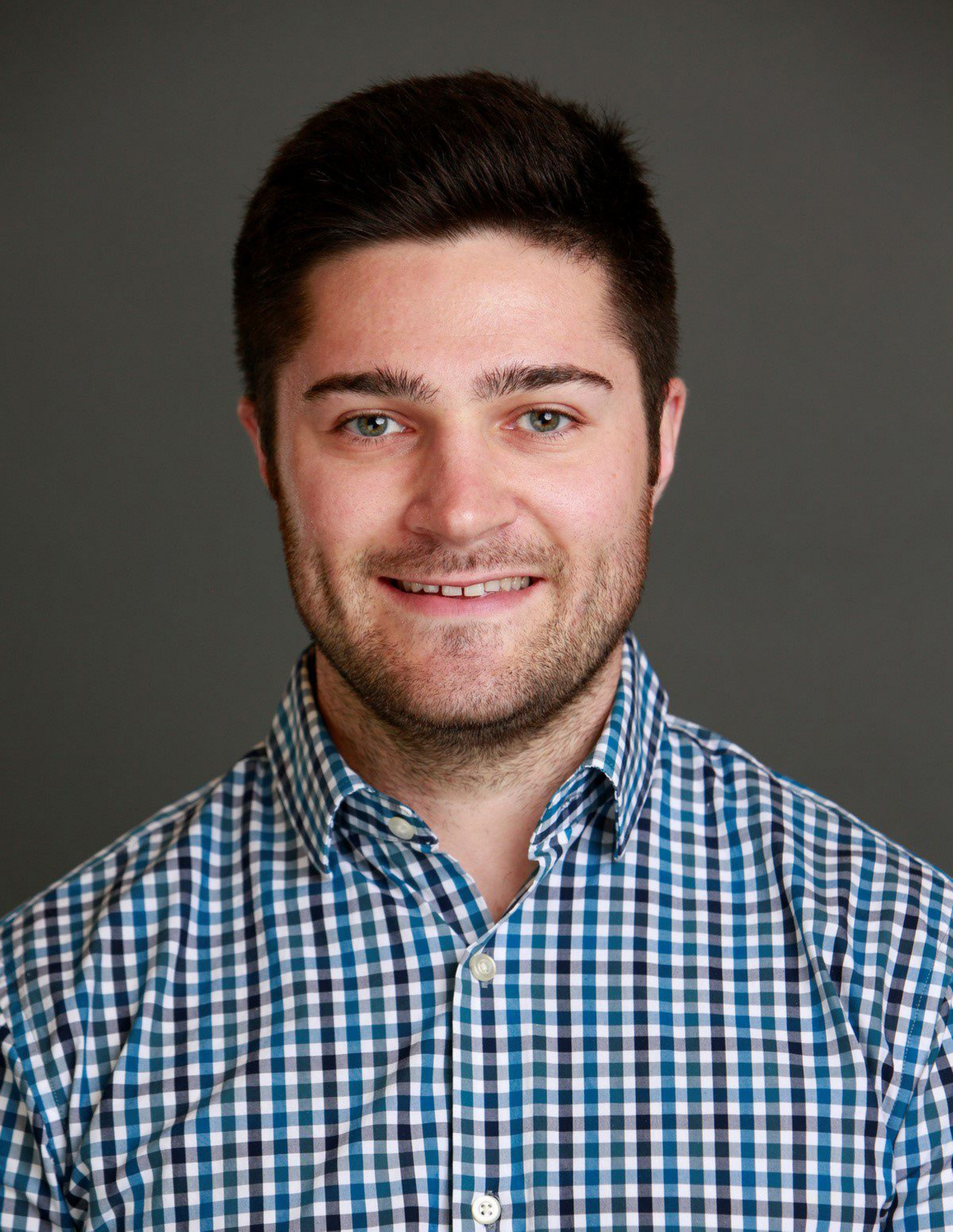}}]{Samuel C.~Chevalier} (Member, IEEE) received M.S. (2016) and B.S. (2015) degrees in Electrical Engineering from the University of Vermont (UVM), and he received the Ph.D. in Mechanical Engineering from the Massachusetts Institute of Technology (MIT) in 2021. During the PhD, his research focused on posing and solving a variety of emerging inverse problems in power systems. He is currently a postdoctoral researcher at the Technical University of Denmark (DTU), where he works on applying learning and optimization to a variety of industry-relevant problems related to the data-driven operation, control, and planning of stochastic power and energy systems.
\end{IEEEbiography}

\begin{IEEEbiography}[{\includegraphics[width=1in,height=1.25in,clip,keepaspectratio]{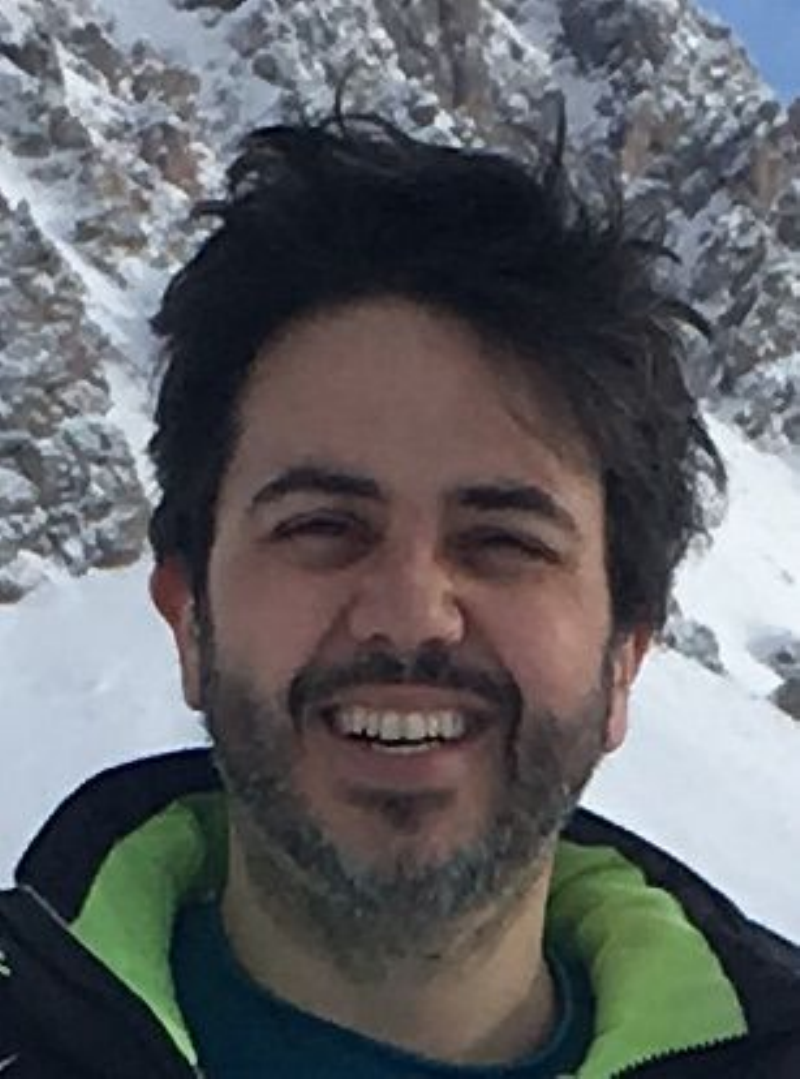}}]{Luca Schenato} (Fellow, IEEE) received the Dr. Eng. degree in electrical engineering from the University of Padova in 1999 and the Ph.D. degree in Electrical Engineering and Computer Sciences from the UC Berkeley, in 2003. He held a post-doctoral position in 2004 and a visiting professor position in 2013-2014 at U.C. Berkeley. Currently he is Full Professor at the Information Engineering Department at the University of Padova. His interests include networked control systems, multi-agent systems, wireless sensor networks, smart grids and cooperative robotics. Luca Schenato has been awarded the 2004 Researchers Mobility Fellowship by the Italian Ministry of Education, University and Research (MIUR), the 2006 Eli Jury Award in U.C. Berkeley and the EUCA European Control Award in 2014, and IEEE Fellow in 2017. He served as Associate Editor for IEEE Trans. on Automatic Control from 2010 to 2014 and he is he is currently Senior Editor for IEEE Trans. on Control of Network Systems and Associate Editor for Automatica.
\end{IEEEbiography} 

\begin{IEEEbiography}[{\includegraphics[width=1in,height=1.25in,clip,keepaspectratio]{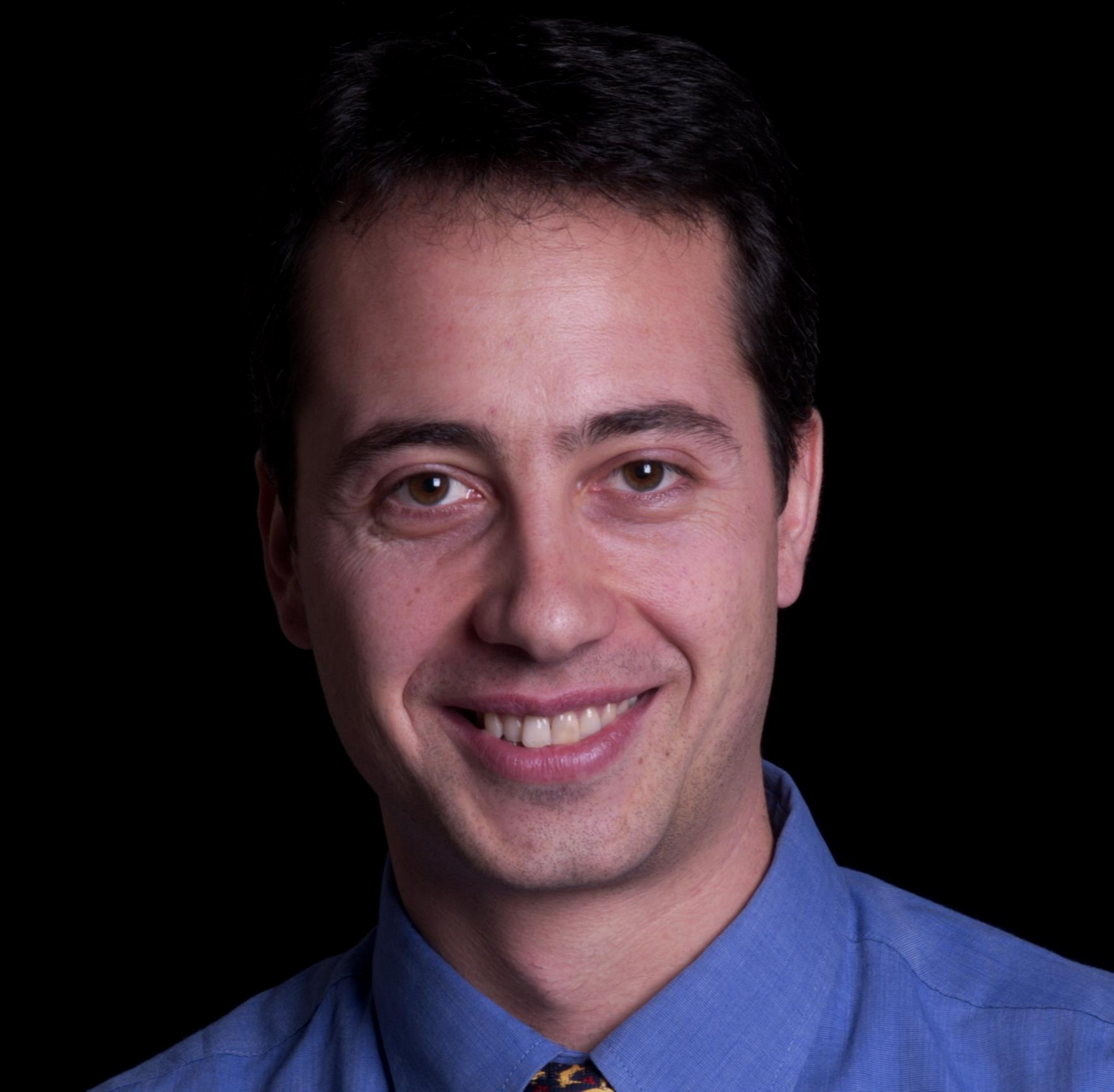}}]{Luca Daniel} (Member, IEEE) is a Professor of Electrical Engineering and Computer Science at the Massachusetts Institute of Technology. His research interests include development of numerical techniques related to uncertainty quantification, inverse problems, robust optimization, parameterized model order reduction and integral equation solvers. His current applications of interest include evaluating and improving robustness of deep neural networks as well as of magnetic resonance imaging scanners, silicon photonics integrated systems, and electrical power distribution networks. Prof. Daniel has received best-paper awards from several journals of the Institutes of Electrical and Electronics Engineers (IEEE), including Transactions on Power Electronics, Transactions on Computer Aided Design, and Transactions on Components and Manufacturing. He has further received 14 best-paper awards at international conferences. Other honors include the IBM Corporation Faculty Award, the IEEE Early Career Award in Electronic Design Automation, and the Spira Award for Excellence in Teaching from the MIT School of Engineering. Dr. Daniel received best PhD thesis awards from both the Department of Electrical Engineering and Computer Sciences and the Department of Applied Mathematics at the University of California at Berkeley, as well as the Outstanding PhD Dissertation Award in Electronic Design Automation from the Association for Computing Machinery (ACM).

\end{IEEEbiography}

\end{document}